\documentstyle[12pt,epsfig]{article} 

\newcommand{\agt}{\,\rlap{\lower 3.5 pt \hbox{$\mathchar \sim$}} \raise 1pt
 \hbox {$>$}\,}
\newcommand{\alt}{\,\rlap{\lower 3.5 pt \hbox{$\mathchar \sim$}} \raise 1pt
 \hbox {$<$}\,}

\catcode`@=11
\newcount\@tempcntc
\def\@citex[#1]#2{\if@filesw\immediate\write\@auxout{\string\citation{#2}}\fi
  \@tempcnta\z@\@tempcntb\m@ne\def\@citea{}\@cite{\@for\@citeb:=#2\do
    {\@ifundefined
       {b@\@citeb}{\@citeo\@tempcntb\m@ne\@citea\def\@citea{,}{\bf ?}\@warning
       {Citation `\@citeb' on page \thepage \space undefined}}%
    {\setbox\z@\hbox{\global\@tempcntc0\csname b@\@citeb\endcsname\relax}%
     \ifnum\@tempcntc=\z@ \@citeo\@tempcntb\m@ne
       \@citea\def\@citea{,}\hbox{\csname b@\@citeb\endcsname}%
     \else
      \advance\@tempcntb\@ne
      \ifnum\@tempcntb=\@tempcntc
      \else\advance\@tempcntb\m@ne\@citeo
      \@tempcnta\@tempcntc\@tempcntb\@tempcntc\fi\fi}}\@citeo}{#1}}
\def\@citeo{\ifnum\@tempcnta>\@tempcntb\else\@citea\def\@citea{,}%
  \ifnum\@tempcnta=\@tempcntb\the\@tempcnta\else
   {\advance\@tempcnta\@ne\ifnum\@tempcnta=\@tempcntb \else \def\@citea{--}\fi
    \advance\@tempcnta\m@ne\the\@tempcnta\@citea\the\@tempcntb}\fi\fi}
\catcode`@=12

\begin{document}

\title{\vskip-3cm{\baselineskip14pt
\centerline{\normalsize DESY 01-015\hfill ISSN 0418-9833}
\centerline{\normalsize hep-ph/0103018\hfill}
\centerline{\normalsize February 2001\hfill}
}
\vskip1.5cm
Pair production of neutral Higgs bosons at the CERN Large Hadron Collider}
\author{A. A. Barrientos Bendez\'u and B. A. Kniehl\\
{\normalsize II. Institut f\"ur Theoretische Physik, Universit\"at Hamburg,}\\
{\normalsize Luruper Chaussee 149, 22761 Hamburg, Germany}}

\date{}

\maketitle

\thispagestyle{empty}

\begin{abstract}
We study the hadroproduction of two neutral Higgs bosons in the minimal 
supersymmetric extension of the standard model (MSSM), which provides a handle
on the trilinear Higgs couplings.
We include the contributions from quark-antiquark annihilation at the tree 
level and those from gluon-gluon fusion, which proceeds via quark and squark 
loops.
We list compact results for the tree-level partonic cross sections and the
squark loop amplitudes, and we confirm previous results for the quark loop 
amplitudes.
We quantitatively analyze the hadronic cross sections at the CERN Large Hadron
Collider assuming a favorable supergravity-inspired MSSM scenario.

\medskip

\end{abstract}

\newpage

\section{Introduction}

The search for Higgs bosons will be among the prime tasks of the CERN Large
Hadron Collider (LHC) \cite{kun}.
While the standard model (SM) of elementary-particle physics contains one
complex Higgs doublet, from which one neutral $CP$-even Higgs boson $H$
emerges in the physical particle spectrum after the spontaneous breakdown of
the electroweak symmetry, the Higgs sector of the minimal supersymmetric
extension of the SM (MSSM) consists of a two-Higgs-doublet model (2HDM) and
accommodates five physical Higgs bosons: the neutral $CP$-even $h^0$ and $H^0$
bosons, the neutral $CP$-odd $A^0$ boson, and the charged $H^\pm$-boson pair.
At the tree level, the MSSM Higgs sector has two free parameters, which are
usually taken to be the mass $m_{A^0}$ of the $A^0$ boson and the ratio
$\tan\beta=v_2/v_1$ of the vacuum expectation values of the two Higgs
doublets.

In the SM, the trilinear and quartic Higgs-boson couplings are proportional to
the square of the Higgs-boson mass, and they are uniquely fixed once the
latter is known.
In the MSSM, the various Higgs-boson self-couplings are determined by the
gauge couplings multiplied with trigonometric factors that depend on $\alpha$,
the mixing angle that rotates the weak $CP$-even Higgs eigenstates into the
mass eigenstates $h^0$ and $H^0$, and $\beta$ \cite{hab} [see 
Eq.~(\ref{eq:hhh})].

At the LHC, the trilinear Higgs-boson couplings may be probed by studying the
inclusive hadroproduction of Higgs-boson pairs.
These couplings enter the stage via the Feynman diagrams where a virtual
neutral Higgs boson is produced in the $s$ channel through $q\bar q$
annihilation or $gg$ fusion and in turn decays into a pair of neutral or 
charged Higgs bosons.
Specifically, the partonic subprocesses include $q\bar q\to HH$ and $gg\to HH$
in the SM, and $q\bar q\to\phi_1\phi_2,H^+H^-$ and $gg\to\phi_1\phi_2,H^+H^-$
in the MSSM, where $\phi_i=h^0,H^0,A^0$.
At the tree level, there are two mechanisms of $q\bar q$ annihilation.
On the one hand, it can proceed via a $Z$ boson (Drell-Yan process) if an
appropriate Higgs-Higgs-$Z$ coupling exists.
On the other hand, the Higgs bosons can be radiated off the $q$-quark line if
the relevant Yukawa couplings are sufficiently strong.
In the SM, $q\bar q$ annihilation is greatly suppressed due to the absence
of a $HHZ$ coupling and the smallness of the $Hq\bar q$ couplings for the 
active quarks contained inside the proton, $q=u,d,s,c,b$.
In the MSSM, however, we have $h^0A^0Z$ and $H^0A^0Z$ couplings at the tree 
level [see Eq.~(\ref{eq:hAZ})], and the $\phi_ib\bar b$ couplings are
generally strong if $\tan\beta$ is large [see Eq.~(\ref{eq:hqq})].
In the SM, $gg$ fusion is mediated via heavy-quark loops.
In the MSSM, there are additional contributions from squark loops.
The SM case was studied in Ref.~\cite{dic}.
As for $H^+H^-$ pair production, $q\bar q$ annihilation was investigated in 
Refs.~\cite{eic,hh}, the quark loop contribution to $gg$ fusion in 
Refs.~\cite{hh,wil,bre}, and the squark loop one in Refs.~\cite{hh,bre}.
As for the pair production of neutral Higgs bosons in the MSSM, $q\bar q$
annihilation via a $Z$ boson was analyzed in Ref.~\cite{daw} and the quark and
squark loop contributions to $gg$ fusion in Refs.~\cite{ple} and \cite{bel},
respectively.
In Ref.~\cite{daw}, also the QCD corrections to $q\bar q$ annihilation via a
$Z$ boson and to $gg$ fusion via an infinitely heavy top quark were 
considered.
The processes of associated production of a neutral Higgs-boson pair with a
dijet (in addition to the remnant jets), an intermediate boson, or another
neutral Higgs boson were found to have cross sections that are greatly
suppressed, by more than an order of magnitude, compared to those of the 
respective processes without those additional final-state particles \cite{kil}.

\begin{figure}[ht]
\begin{center}
\begin{tabular}{cc}
\parbox{8cm}{\epsfig{figure=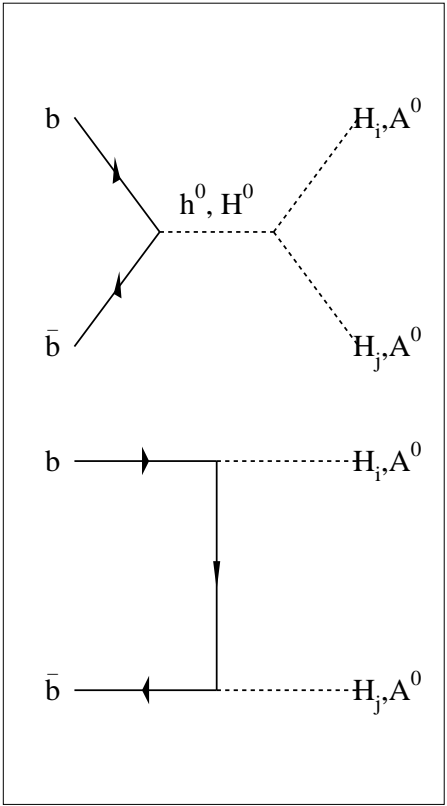,width=8cm}} &
\parbox{8cm}{\epsfig{figure=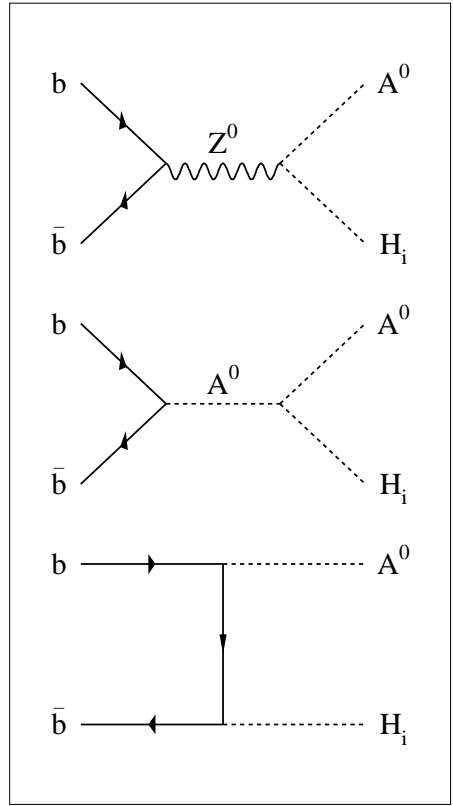,width=8cm}}\\
(a) & (b)
\end{tabular}
\caption{Tree-level Feynman diagrams for $q\bar q\to\phi_1\phi_2$, with (a)
$\phi_1\phi_2=h^0h^0,h^0H^0,H^0H^0,A^0A^0$ and (b)
$\phi_1\phi_2=h^0A^0,H^0A^0$, in the MSSM.}
\label{fig:tree}
\end{center}
\end{figure}

The purpose of this paper is to reanalyze the pair production of neutral Higgs 
bosons in the MSSM, both via $q\bar q$ annihilation and $gg$ fusion.
In the case of $q\bar q$ annihilation, we also allow for $q=b$.
This makes it necessary to include a new class of diagrams involving $b$-quark
Yukawa couplings, which are depicted in Fig.~\ref{fig:tree}(a) and the second
and third lines of Fig.~\ref{fig:tree}(b).
These come in addition to the Drell-Yan diagram, shown in the first line of
Fig.~\ref{fig:tree}(b), which is already present for $q=u,d,s,c$.
For final states with an even number of $CP$-even Higgs bosons,
$\phi_1\phi_2=h^0h^0,h^0H^0,H^0H^0,A^0A^0$, $b\bar b$ annihilation is the only
production mechanism at the tree level.
As we shall see, its cross section is comparable to --- and in certain areas
of the MSSM parameter space even in excess of --- the one of $gg$ fusion.
For final states with an odd number of $CP$-even Higgs bosons,
$\phi_1\phi_2=h^0A^0,H^0A^0$, $b\bar b$ annihilation significantly enhances
the Drell-Yan contribution for large values of $\tan\beta$.
As for $gg$ fusion, we reproduce the analytical and numerical results for the
quark loop contributions of Ref.~\cite{ple}.
Apart from obvious typographical errors, we also agree with the formulas for
the squark loop contributions listed in Ref.~\cite{bel}.
However, we find their numerical size to be considerably smaller than what was
found in Ref.~\cite{bel}.

As for $b\bar b$ annihilation, it should be noted that the treatment of bottom
as an active flavor inside the colliding hadrons leads to an effective
description, which comprises contributions from the higher-order subprocesses
$gb\to\phi_1\phi_2b$, $g\bar b\to\phi_1\phi_2\bar b$, and
$gg\to\phi_1\phi_2b\bar b$.
If all these subprocesses are to be explicitly included along with
$b\bar b\to\phi_1\phi_2$, then it is necessary to employ a judiciously
subtracted parton density function (PDF) for the $b$ quark in order to avoid
double counting \cite{gun}.
The evaluation of $b\bar b\to\phi_1\phi_2$ with an unsubtracted $b$-quark PDF
is expected to slightly overestimate the true cross section \cite{gun}.
For simplicity, we shall nevertheless adopt this effective approach in our
analysis, keeping in mind that a QCD-correction factor below unity is to be 
applied.

In order to reduce the number of unknown supersymmetric input parameters, we 
adopt a scenario where the MSSM is embedded in a grand unified theory (GUT)
involving supergravity (SUGRA) \cite{kal}.
The MSSM thus constrained is characterized by the following parameters at the
GUT scale, which come in addition to $\tan\beta$ and $m_{A^0}$: the universal
scalar mass $m_0$, the universal gaugino mass $m_{1/2}$, the trilinear
Higgs-sfermion coupling $A$, the bilinear Higgs coupling $B$, and the
Higgs-higgsino mass parameter $\mu$.
Notice that $m_{A^0}$ is then not an independent parameter anymore, but it is
fixed through the renormalization group equation.
The number of parameters can be further reduced by making additional
assumptions.
Unification of the $\tau$-lepton and $b$-quark Yukawa couplings at the GUT
scale leads to a correlation between $m_t$ and $\tan\beta$.
Furthermore, if the electroweak symmetry is broken radiatively, then $B$ and
$\mu$ are determined up to the sign of $\mu$.
Finally, it turns out that the MSSM parameters are nearly independent of the
value of $A$, as long as $|A|\alt500$~GeV at the GUT scale.

This paper is organized as follows.
In Sec.~\ref{sec:two}, we list analytic results for the tree-level cross
sections of $q\bar q\to\phi_1\phi_2$, including the Yukawa-enhanced
contributions for $q=b$, and the squark loop contributions to the
$gg\to\phi_1\phi_2$ amplitudes in the MSSM.
The relevant MSSM coupling constants and the squark loop form factors are
relegated to Appendices~A and B, respectively.
In Sec.~\ref{sec:three}, we present quantitative predictions for the inclusive
cross section of $pp\to\phi_1\phi_2+X$ at the LHC adopting a favorable
SUGRA-inspired MSSM scenario.
Sec.~\ref{sec:four} contains our conclusions.

\section{\label{sec:two}Analytic Results}

In this section, we present the tree-level cross sections of the partonic
subprocesses $q\bar q\to\phi_1\phi_2$, where $\phi_i=h^0,H^0,A^0$, and the
transition ($T$) matrix elements of $gg\to\phi_1\phi_2$ arising from squark
triangle and box diagrams.

We work in the parton model of QCD with $n_f=5$ active quark flavors, which we
take to be massless.
However, we retain the $b$-quark Yukawa couplings at their finite values, in 
order not to suppress possibly sizeable contributions.
We adopt the MSSM Feynman rules from Ref.~\cite{hab}.
In Appendix~A, we list the trilinear self-couplings of the neutral Higgs
bosons and their couplings to gauge bosons and quarks.
For each quark flavor $q$ there is a corresponding squark flavor $\tilde q$,
which comes in two mass eigenstates $i=1,2$.
The masses $m_{\tilde q_i}$ of the squarks and their trilinear couplings to
the $h^0$ and $H^0$ bosons are listed in Eq.~(A.5) of
Ref.~\cite{hem}\footnote{In Ref.~\cite{hem}, $m_{\tilde q_i}$ is called
$M_{\tilde Qa}$.} and Eq.~(A.2) of Ref.~\cite{hh}, respectively.
Their trilinear couplings to the $A^0$ boson and their quartic couplings to
the $h^0$, $H^0$, and $A^0$ bosons may be found in Appendix~A.

Considering the generic partonic subprocess $ab\to\phi_1\phi_2$, we denote the
four-momenta of the incoming partons, $a$ and $b$, and the outgoing Higgs
bosons, $\phi_1$ and $\phi_2$, by $p_a$, $p_b$, $p_1$, and $p_2$,
respectively, and define the partonic Mandelstam variables as $s=(p_a+p_b)^2$,
$t=(p_a-p_1)^2$, and $u=(p_b-p_1)^2$.
The on-shell conditions read $p_a^2=p_b^2=0$ and $p_i^2=h_i$, where $h_i$
denotes the square of the $\phi_i$-boson mass.
Four-momentum conservation implies that $s+t+u=h_1+h_2$.
Furthermore, we have $sp_T^2=tu-h_1h_2$, where $p_T$ is the absolute value of
transverse momentum common to $\phi_1$ and $\phi_2$ in the center-of-mass
(c.m.) frame.

The tree-level diagrams for $b\bar b\to\phi_1\phi_2$, with
$\phi_1\phi_2=h^0h^0,h^0H^0,H^0H^0,A^0A^0$ and $\phi_1\phi_2=h^0A^0,H^0A^0$,
are depicted in Figs.~\ref{fig:tree}(a) and (b), respectively.
The cross sections for the first class of partonic subprocesses may be 
generically written as
\begin{equation}
\frac{d\sigma}{dt}(b\bar b\to\phi_1\phi_2)=\frac{1}{1+\delta_{\phi_1\phi_2}}\,
\frac{G_F^2m_W^4}{3\pi s}\left(|S|^2+p_T^2T_-^2\right),
\label{eq:hh}
\end{equation}
where the prefactor accounts for identical-particle symmetrization if
$\phi_1=\phi_2$, $G_F$ is Fermi's constant, $m_W$ is the $W$-boson mass,
\begin{eqnarray}
S&=&g_{\phi_1\phi_2h^0}g_{h^0bb}{\cal P}_{h^0}(s)
+g_{\phi_1\phi_2H^0}g_{H^0bb}{\cal P}_{H^0}(s),
\nonumber\\
T_\pm&=&g_{\phi_1bb}g_{\phi_2bb}\left(\frac{1}{t}\pm\frac{1}{u}\right).
\end{eqnarray}
Here,
\begin{equation}
{\cal P}_{X}(s)=\frac{1}{s-m_X^2+im_X\Gamma_X}
\end{equation}
is the propagator function of particle $X$, with mass $m_X$ and total decay
width $\Gamma_X$.
For the second class of partonic subprocesses, we have
\begin{equation}
\frac{d\sigma}{dt}(b\bar b\to\phi_1\phi_2)=
\frac{G_F^2m_W^4}{3\pi s}\left[|P|^2+p_T^2(|V|^2+|A-T_+|^2)\right],
\label{eq:hA}
\end{equation}
where
\begin{eqnarray}
P&=&g_{\phi_1\phi_2A^0}g_{A^0bb}{\cal P}_{A^0}(s),
\nonumber\\
V&=&2g_{\phi_1\phi_2Z}v_{Zbb}{\cal P}_{Z}(s),\nonumber\\
A&=&2g_{\phi_1\phi_2Z}a_{Zbb}{\cal P}_{Z}(s).
\end{eqnarray}
Here, $v_{Zbb}=-\left(I_b-2s_w^2Q_b\right)/(2c_w)$ and $a_{Zbb}=-I_b/(2c_w)$,
with $c_w^2=1-s_w^2=m_W^2/m_Z^2$, are the vector and axial-vector couplings of
the $b$ quark, with weak isospin $I_b=-1/2$ and electric charge $Q_b=-1/3$, to
the $Z$ boson.
As for $h^0A^0$ and $H^0A^0$ production, there are also sizeable contributions
from $q\bar q$ annihilation via a $Z$ boson for the quarks of the first and
second generations, $q=u,d,s,c$.
The corresponding Drell-Yan cross sections are obtained from Eq.~(\ref{eq:hA})
by putting $P=T_+=0$ and substituting $b\to q$.
The resulting expression agrees with Eq.~(36) of Ref.~\cite{daw}.
The full tree-level cross sections are then obtained by complementing the
$b\bar b$-initiated cross sections of Eq.~(\ref{eq:hA}) with the Drell-Yan
cross sections for $q=u,d,s,c$.

\begin{figure}[ht]
\begin{center}
\begin{tabular}{cc}
\parbox{8cm}{\epsfig{figure=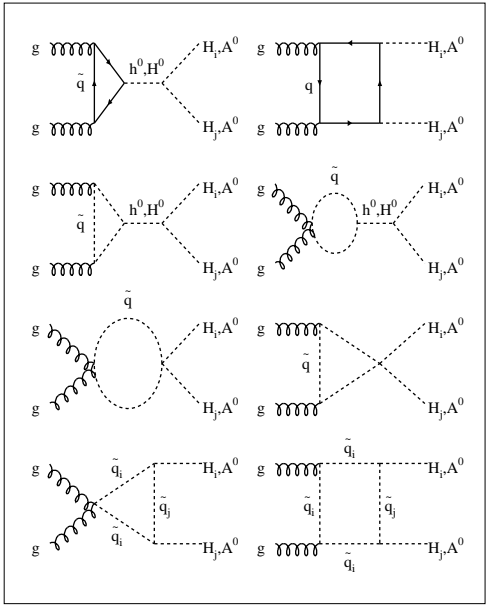,width=8cm}} &
\parbox{8cm}{\epsfig{figure=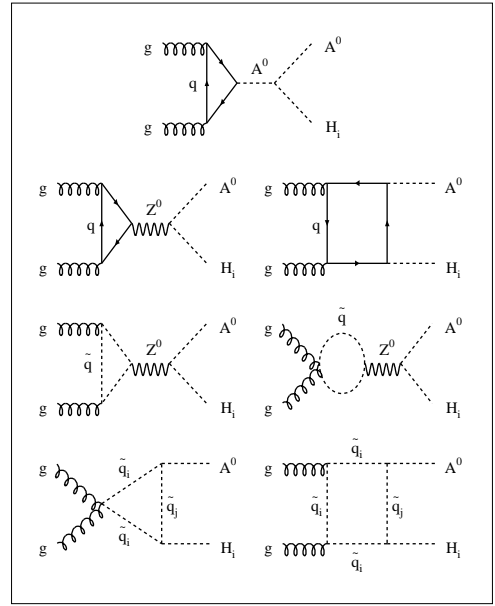,width=8cm}} \\
(a) & (b)
\end{tabular}
\caption{One-loop Feynman diagrams for $gg\to\phi_1\phi_2$, with (a)
$\phi_1\phi_2=h^0h^0,h^0H^0,H^0H^0,A^0A^0$ and (b) 
$\phi_1\phi_2=h^0A^0,H^0A^0$, due to virtual quarks and squarks in the MSSM.}
\label{fig:loop}
\end{center}
\end{figure}

The one-loop diagrams for $gg\to\phi_1\phi_2$, with
$\phi_1\phi_2=h^0h^0,h^0H^0,H^0H^0,A^0A^0$ and $\phi_1\phi_2=h^0A^0,H^0A^0$,
are depicted in Figs.~\ref{fig:loop}(a) and (b), respectively.
As for the quark loops, our analytical results fully agree with those listed
in Ref.~\cite{ple}, and there is no need to repeat them here.
For the partonic subprocesses of class two, the squark loop contributions are
zero \cite{bel}.
This may be understood as follows.
(i) The $g_{g\tilde q_i\tilde q_j}$ and $g_{Z\tilde q_i\tilde q_j}$ couplings
are linear in the squark four-momenta, while the $g_{gg\tilde q_i\tilde q_j}$
couplings are momentum independent.
Thus, the diagrams in the third line of Fig.~\ref{fig:loop}(b) each vanish
upon adding their counterparts with the loop-momentum flows reversed.
(ii) The $g_{g\tilde q_i\tilde q_j}$, $g_{gg\tilde q_i\tilde q_j}$,
$g_{h^0\tilde q_i\tilde q_j}$, and $g_{H^0\tilde q_i\tilde q_j}$ couplings are
symmetric in $i$ and $j$, while the $g_{A^0\tilde q_i\tilde q_j}$ coupling is
antisymmetric.
Thus, the diagrams in the last line of Fig.~\ref{fig:loop}(b) vanish upon
summation over $i$ and $j$.
For the partonic subprocesses of class one, the $T$-matrix elements
corresponding to the squark triangle and box diagrams are found to be
\begin{eqnarray}
\tilde{\cal T}_\triangle&=&\frac{G_Fm_W^2}{\sqrt2}\,
\frac{\alpha_s(\mu_r)}{\pi}
\varepsilon_\mu^c(p_a)\varepsilon_\nu^c(p_b)A_1^{\mu\nu}\tilde F_\triangle,
\nonumber\\
\tilde{\cal T}_\Box&=&\frac{G_Fm_W^2}{\sqrt2}\,\frac{\alpha_s(\mu_r)}{\pi}
\varepsilon_\mu^c(p_a)\varepsilon_\nu^c(p_b)
\left(A_1^{\mu\nu}\tilde F_\Box+A_2^{\mu\nu}\tilde G_\Box\right),
\end{eqnarray}
respectively, where $\alpha_s(\mu_r)$ is the strong-coupling constant at
renormalization scale $\mu_r$, $\varepsilon_\mu^c(p_a)$ is the polarization
four-vector of gluon $a$ and similarly for gluon $b$, it is summed over the
color index $c=1,\ldots,8$,
\begin{eqnarray}
A_1^{\mu\nu}&=&g^{\mu\nu}-\frac{2}{s}p_b^\mu p_a^\nu,
\nonumber\\
A_2^{\mu\nu}&=&g^{\mu\nu}+\frac{2}{p_T^2}\left(\frac{h_1}{s}p_b^\mu p_a^\nu
+\frac{u-h_1}{s}p_1^\mu p_a^\nu+\frac{t-h_1}{s}p_b^\mu p_1^\nu
+p_1^\mu p_1^\nu\right),
\end{eqnarray}
and the form factors $\tilde F_\triangle$, $\tilde F_\Box$, and
$\tilde G_\Box$ are listed in Appendix~B.
Due to Bose symmetry, $\tilde{\cal T}_\triangle$ and $\tilde{\cal T}_\Box$ are
invariant under the simultaneous replacements $\mu\leftrightarrow\nu$ and
$p_a\leftrightarrow p_b$.
Consequently, $\tilde F_\triangle$, $\tilde F_\Box$, and $\tilde G_\Box$ are
symmetric in $t$ and $u$.
Our analytic results for the squark loop contributions agree with those given
in Eqs.~(8)--(10) of Ref.~\cite{bel}, which are expressed in terms of helicity 
amplitudes.\footnote{There are two obvious typographical errors on the
right-hand side of Eq.~(10e) in Ref.~\cite{bel}: There should be an overall
minus sign, and $V_{H_{(i,j)}\tilde q_k\tilde q_k}$ should be replaced by
$V_{H_{(i,j)}\tilde q_k\tilde q_l}$.}

The parton-level cross section of $gg\to\phi_1\phi_2$ including both quark and 
squark contributions is then given by
\begin{eqnarray}
\frac{d\sigma}{dt}(gg\to\phi_1\phi_2)&=&\frac{1}{1+\delta_{\phi_1\phi_2}}\,
\frac{G_F^2\alpha_s^2(\mu_r)}{256(2\pi)^3}
\left[\left|\sum_{Q=t,b}C_\triangle^QF_\triangle^Q+F_\Box
-\frac{2m_W^2}{s}\left(\tilde F_\triangle+\tilde F_\Box\right)\right|^2
\right.\nonumber\\
&&{}+\left.\left|G_\Box-\frac{2m_W^2}{s}\tilde G_\Box\right|^2
+\left|H_\Box\right|^2\right],
\end{eqnarray}
where the generalized couplings $C_\triangle^Q$ and $C_\Box^Q$ and the form
factors $F_\triangle^Q$, $F_\Box$, and $G_\Box$ may be found in Eq.~(16)--(18)
and Appendix~A of Ref.~\cite{ple}, respectively.

The kinematics of the inclusive reaction $AB\to CD+X$, where $A$ and
$B$ are hadrons, which are taken to be massless, and $C$ and $D$ are massive
particles, is described in Sec.~II of Ref.~\cite{wh}.
Its double-differential cross section $d^2\sigma/dy\,dp_T$, where $y$ and
$p_T$ are the rapidity and transverse momentum of particle $C$ in the c.m.\
frame of the hadronic collision, may be evaluated from Eq.~(2.1) of
Ref.~\cite{wh}.

\section{\label{sec:three}Phenomenological Implications}

We are now in a position to explore the phenomenological implications of our
results.
The SM input parameters for our numerical analysis are taken to be
$G_F=1.16639\times10^{-5}$~GeV$^{-2}$, $m_W=80.419$~GeV, $m_Z=91.1882$~GeV,
$m_t=174.3$~GeV , and $m_b=4.6$~GeV \cite{pdg}.
We adopt the lowest-order (LO) proton PDF set CTEQ5L \cite{lai}.
We evaluate $\alpha_s(\mu_r)$ from the LO formula \cite{pdg} with
$n_f=5$ quark flavors and asymptotic scale parameter
$\Lambda_{\rm QCD}^{(5)}=146$~MeV \cite{lai}.
We identify the renormalization and factorization scales with the
$\phi_1\phi_2$ invariant mass $\sqrt s$, $M=\mu_r=\sqrt s$.
We vary $\tan\beta$ and $m_{A^0}$ in the ranges
$3<\tan\beta<38\approx m_t/m_b$ and 90~GeV${}<m_{A^0}<1$~TeV, respectively.
As for the GUT parameters, we choose $m_{1/2}=150$~GeV, $A=0$, and $\mu<0$, 
and tune $m_0$ so as to be consistent with the desired value of $m_{A^0}$.
All other MSSM parameters are then determined according to the SUGRA-inspired
scenario as implemented in the program package SUSPECT \cite{djo}.
For the typical example of $\tan\beta=3$ and $m_{A^0}=300$~GeV, the residual
masses and total decay widths of the $\phi_i$ bosons are
$m_{h^0}=90$~GeV, $m_{H^0}=306$~GeV,
$\Gamma_{h^0}=3$~MeV, $\Gamma_{H^0}=186$~MeV, and $\Gamma_{A^0}=72$~MeV,
and the squark masses are
$m_{\tilde u_1}=m_{\tilde c_1}=412$~GeV,
$m_{\tilde u_2}=m_{\tilde c_2}=422$~GeV,
$m_{\tilde d_1}=m_{\tilde s_1}=413$~GeV,
$m_{\tilde d_2}=m_{\tilde s_2}=428$~GeV,
$m_{\tilde t_1}=317$~GeV, $m_{\tilde t_2}=443$~GeV,
$m_{\tilde b_1}=384$~GeV, and $m_{\tilde b_2}=413$~GeV,
We do not impose the unification of the $\tau$-lepton and $b$-quark Yukawa
couplings at the GUT scale, which would just constrain the allowed $\tan\beta$
range without any visible effect on the results for these values of
$\tan\beta$.
We exclude solutions which do not comply with the present experimental lower
mass bounds of the sfermions, charginos, neutralinos, and Higgs bosons
\cite{ruh}.
In our analysis, an $s$-channel resonance only occurs in the process
$pp\to h^0h^0+X$ if $m_{H^0}>2m_{h^0}$.

\begin{figure}[ht]
\begin{center}
\begin{tabular}{c}
\parbox{12cm}{\epsfig{figure=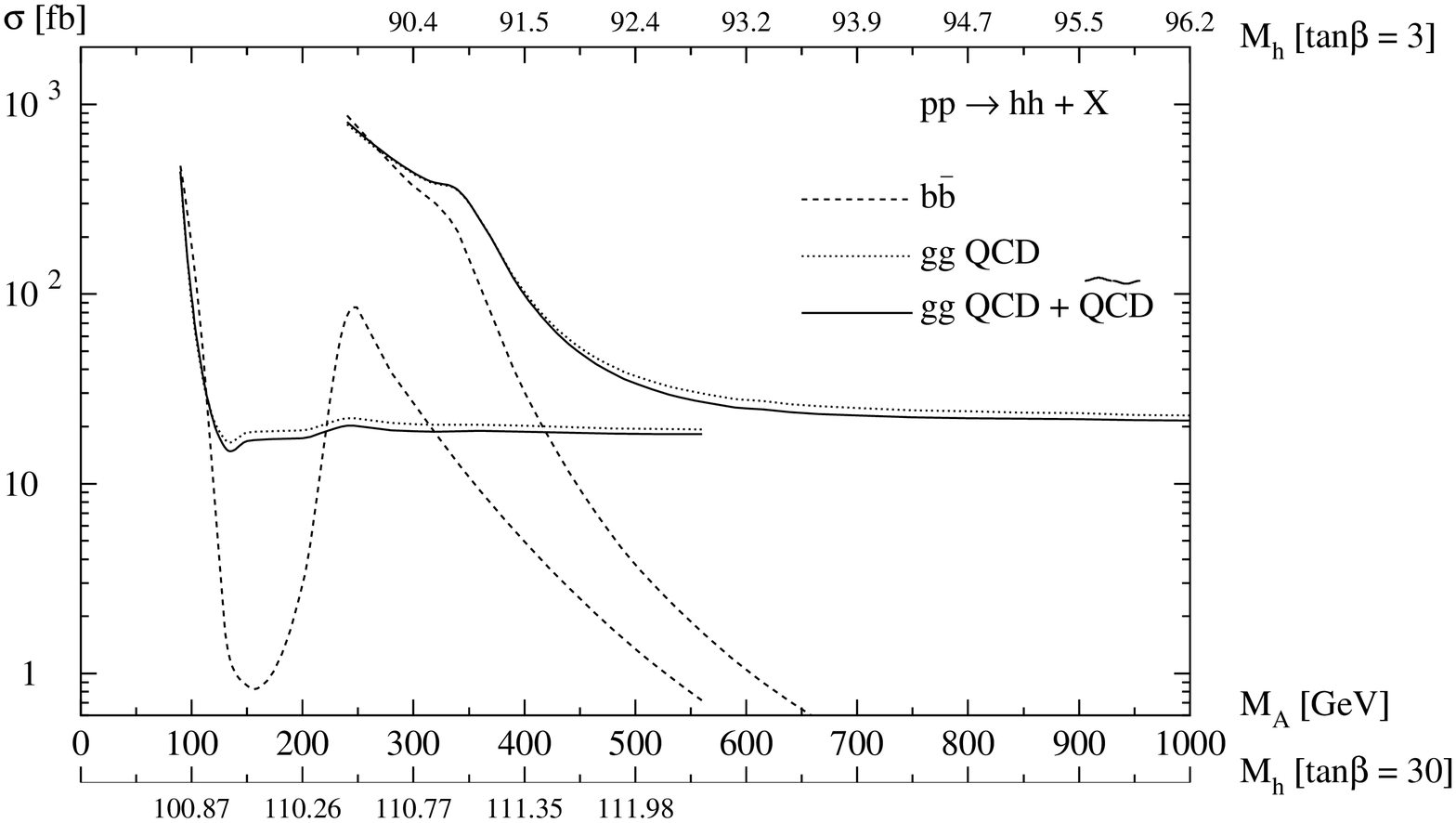,width=12cm}} \\
(a) \\
\parbox{12cm}{\epsfig{figure=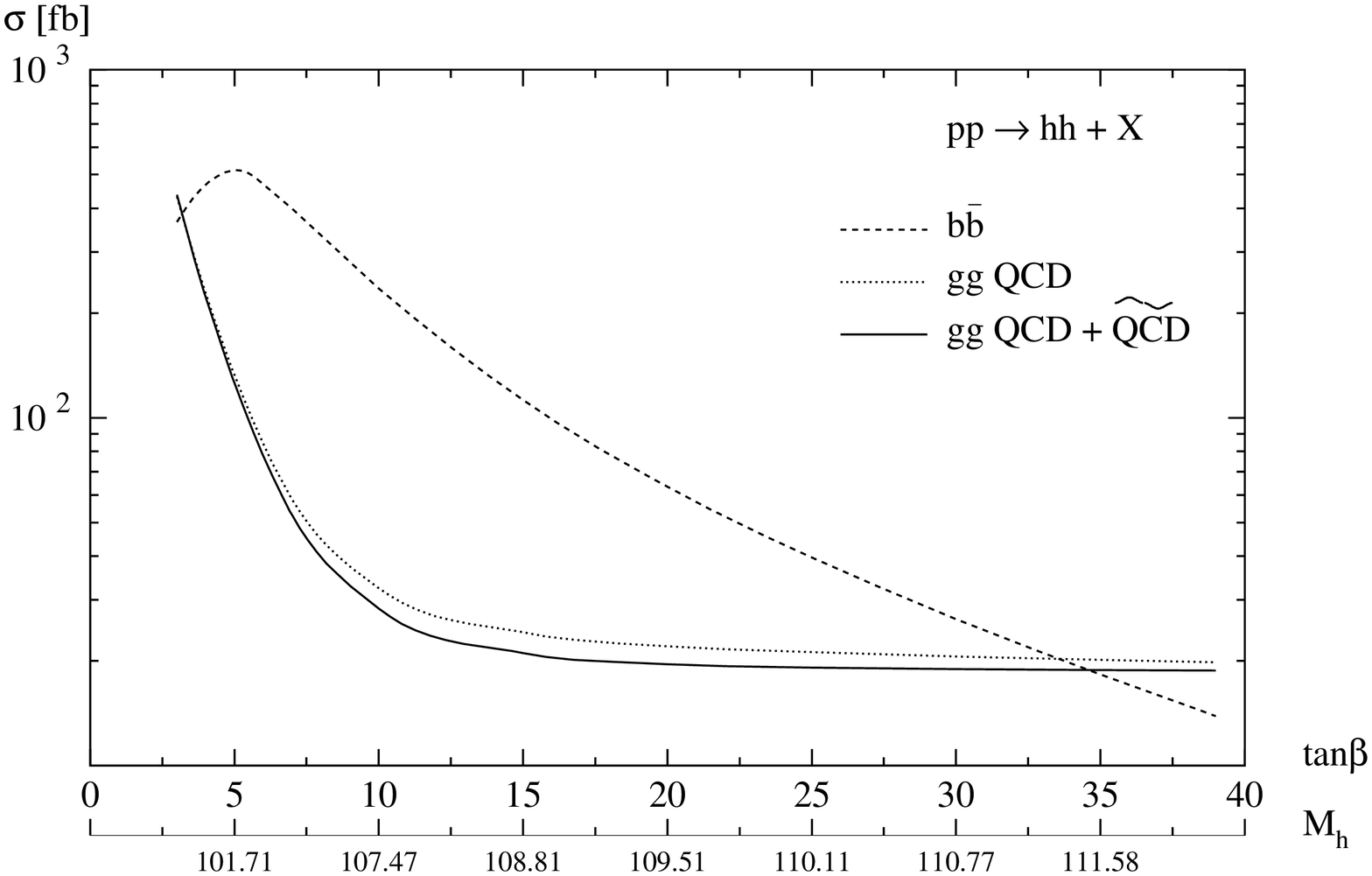,width=12cm}} \\
(b)
\end{tabular}
\caption{Total cross sections $\sigma$ (in fb) of $pp\to h^0h^0+X$ via
$b\bar b$ annihilation (dashed lines) and $gg$ fusion (solid lines) at the LHC
(a) as functions of $m_{A^0}$ for $\tan\beta=3$ (starting at 
$m_{A^0}=240$~GeV) and 30 (starting at $m_{A^0}=90$~GeV); and (b) as functions
of $\tan\beta$ for $m_{A^0}=300$~GeV.
For comparison, also the quark loop contribution to $gg$ fusion (dotted lines)
is shown.}
\label{fig:hh}
\end{center}
\end{figure}

\begin{figure}[ht]
\begin{center}
\begin{tabular}{c}
\parbox{12cm}{\epsfig{figure=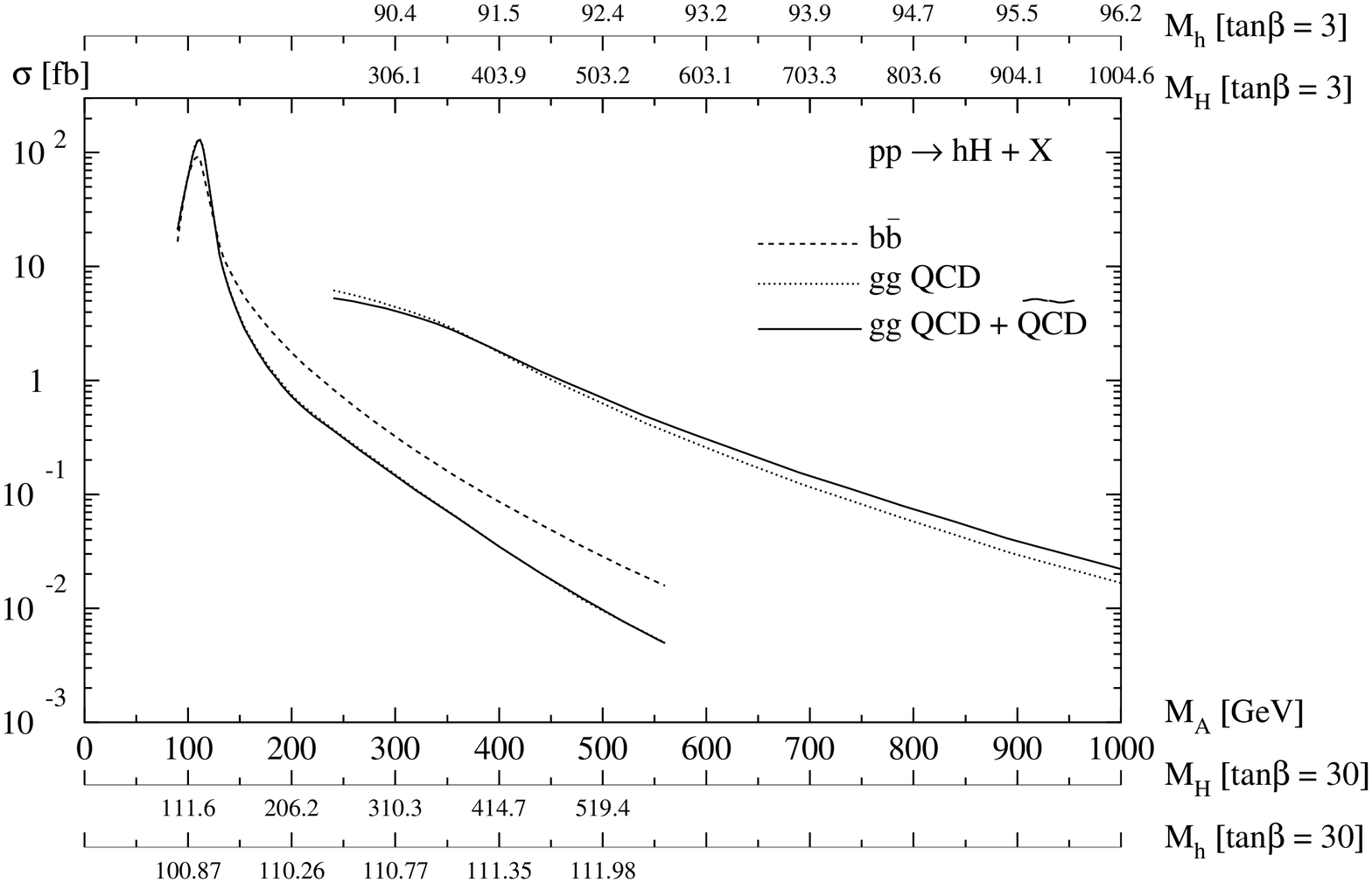,width=12cm}} \\
(a) \\
\parbox{12cm}{\epsfig{figure=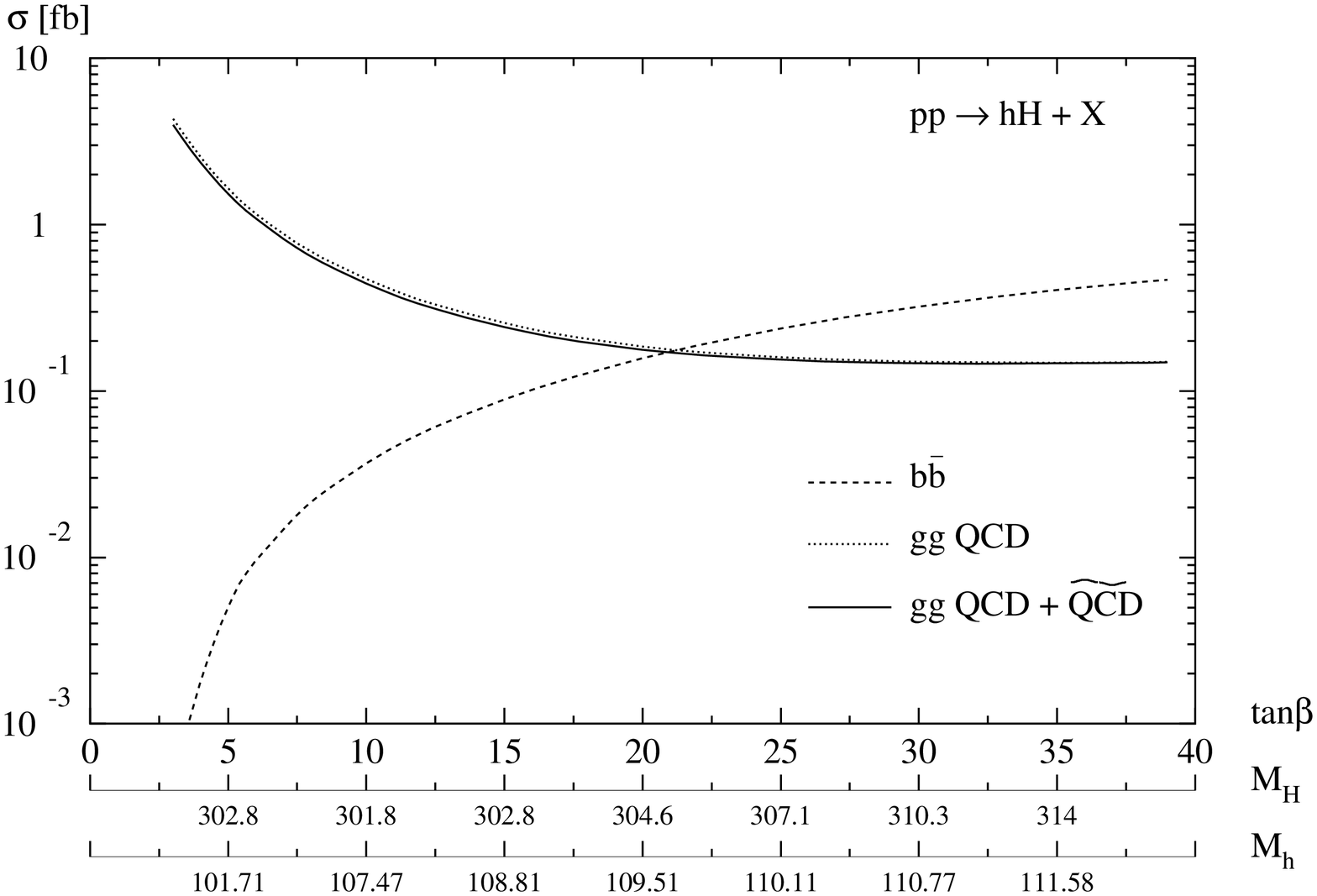,width=12cm}} \\
(b)
\end{tabular}
\caption{Total cross sections $\sigma$ (in fb) of $pp\to h^0H^0+X$ via
$b\bar b$ annihilation (dashed lines) and $gg$ fusion (solid lines) at the LHC
(a) as functions of $m_{A^0}$ for $\tan\beta=3$ (starting at 
$m_{A^0}=240$~GeV) and 30 (starting at $m_{A^0}=90$~GeV); and (b) as functions
of $\tan\beta$ for $m_{A^0}=300$~GeV.
For comparison, also the quark loop contribution to $gg$ fusion (dotted lines)
is shown.}
\label{fig:hH}
\end{center}
\end{figure}

\begin{figure}[ht]
\begin{center}
\begin{tabular}{c}
\parbox{12cm}{\epsfig{figure=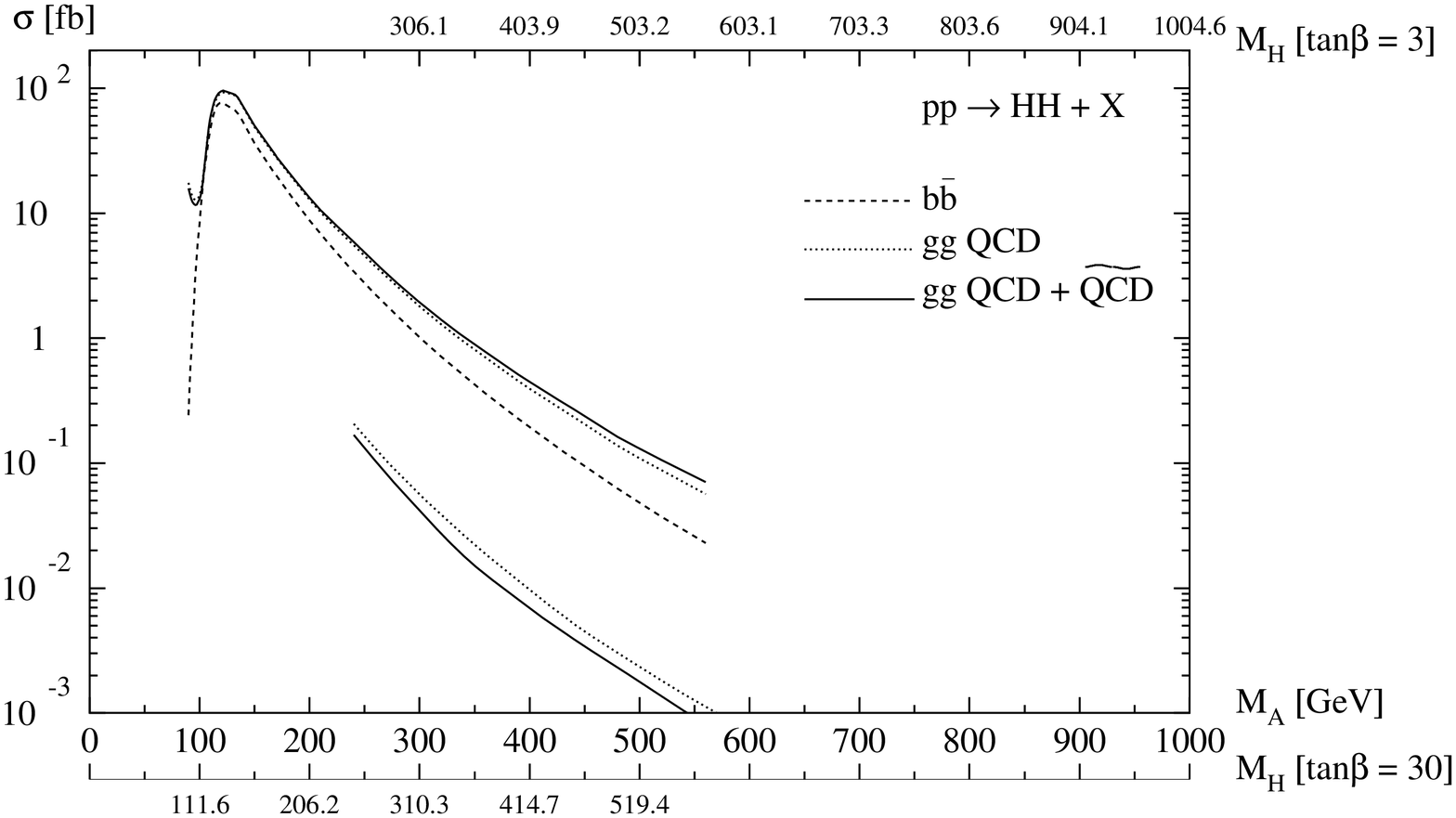,width=12cm}} \\
(a) \\
\parbox{12cm}{\epsfig{figure=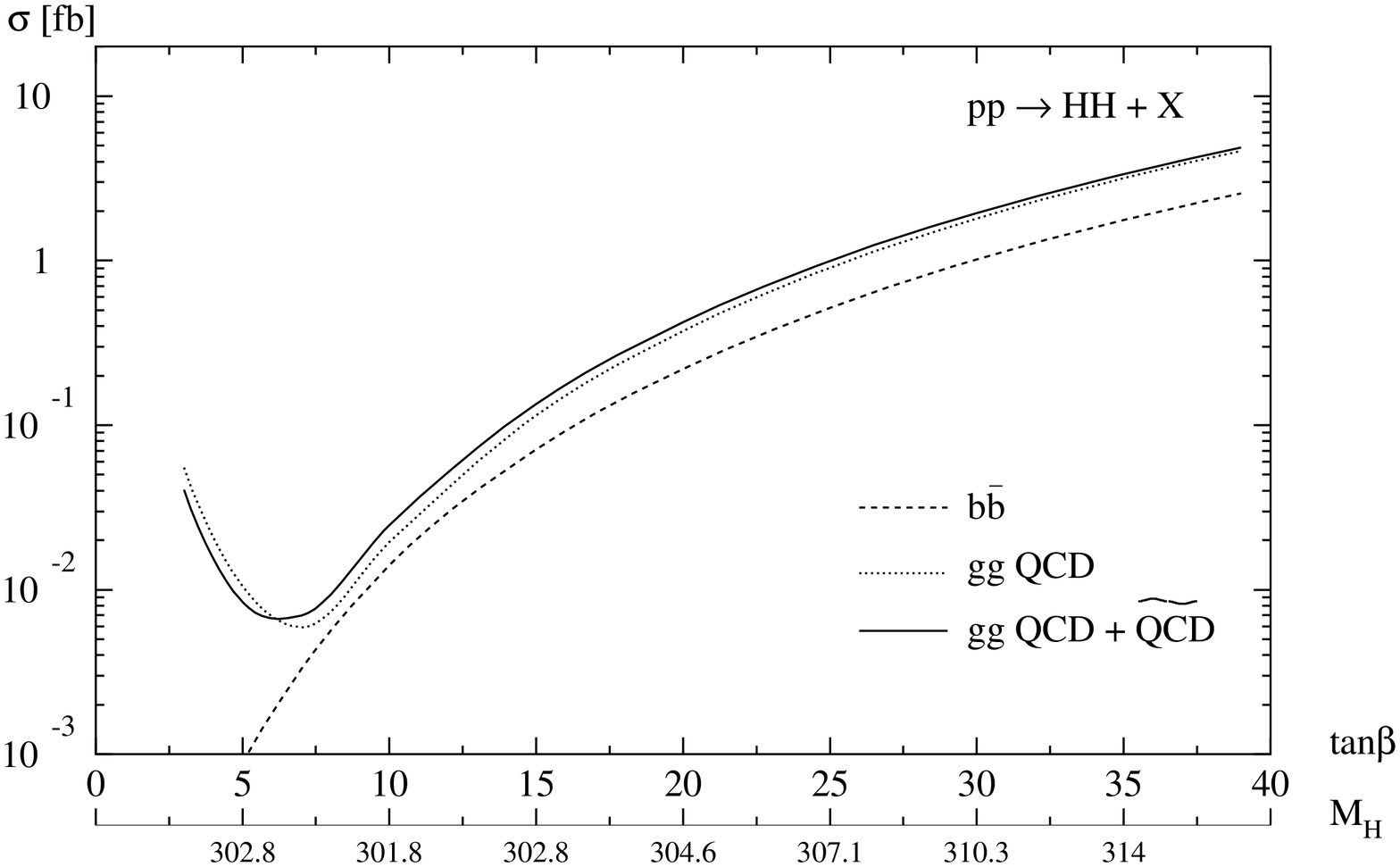,width=12cm}} \\
(b)
\end{tabular}
\caption{Total cross sections $\sigma$ (in fb) of $pp\to H^0H^0+X$ via
$b\bar b$ annihilation (dashed lines) and $gg$ fusion (solid lines) at the LHC
(a) as functions of $m_{A^0}$ for $\tan\beta=3$ (starting at 
$m_{A^0}=240$~GeV) and 30 (starting at $m_{A^0}=90$~GeV); and (b) as functions
of $\tan\beta$ for $m_{A^0}=300$~GeV.
For comparison, also the quark loop contribution to $gg$ fusion (dotted lines)
is shown.}
\label{fig:HH}
\end{center}
\end{figure}

\begin{figure}[ht]
\begin{center}
\begin{tabular}{c}
\parbox{12cm}{\epsfig{figure=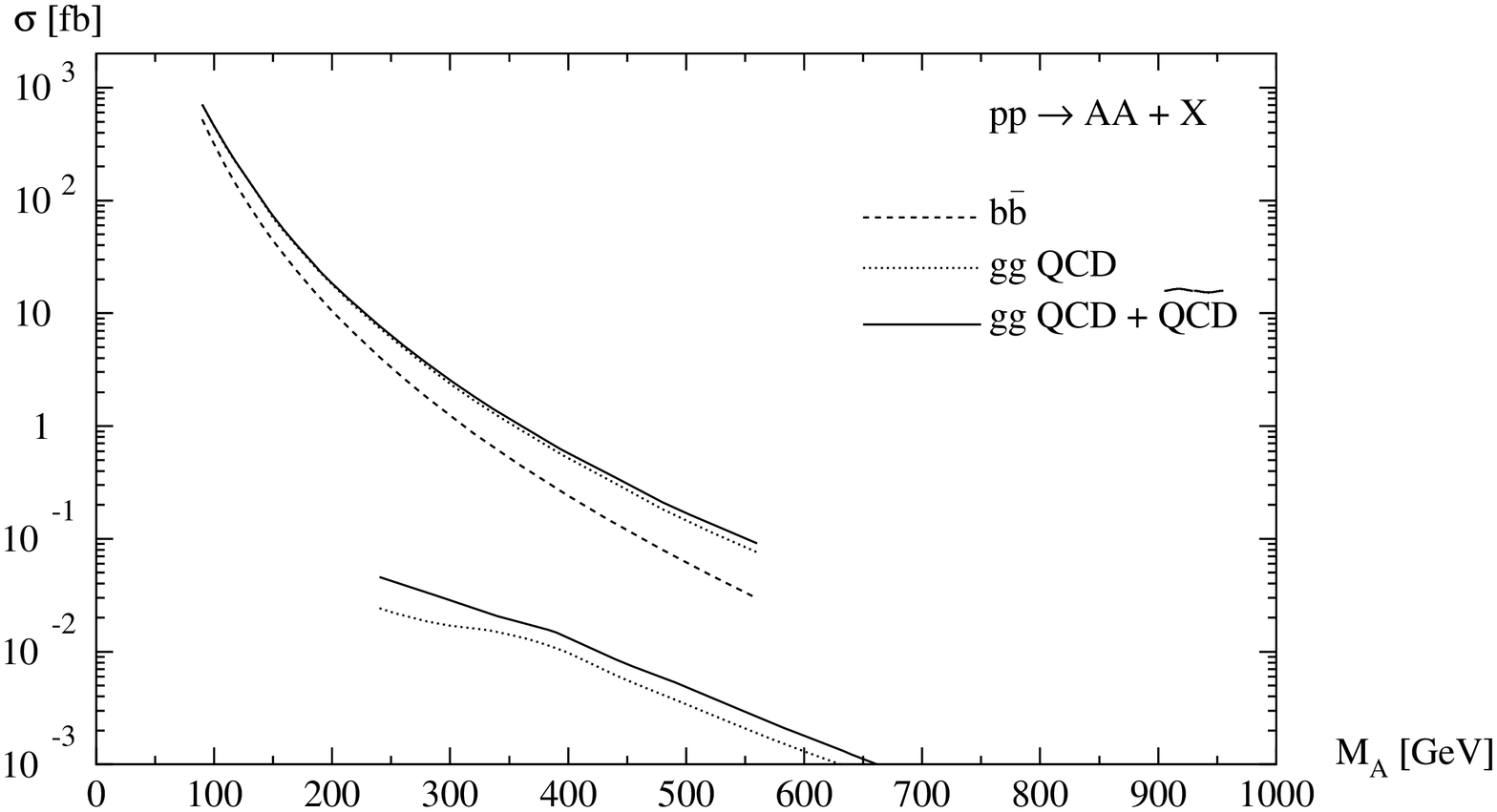,width=12cm}} \\
(a) \\
\parbox{12cm}{\epsfig{figure=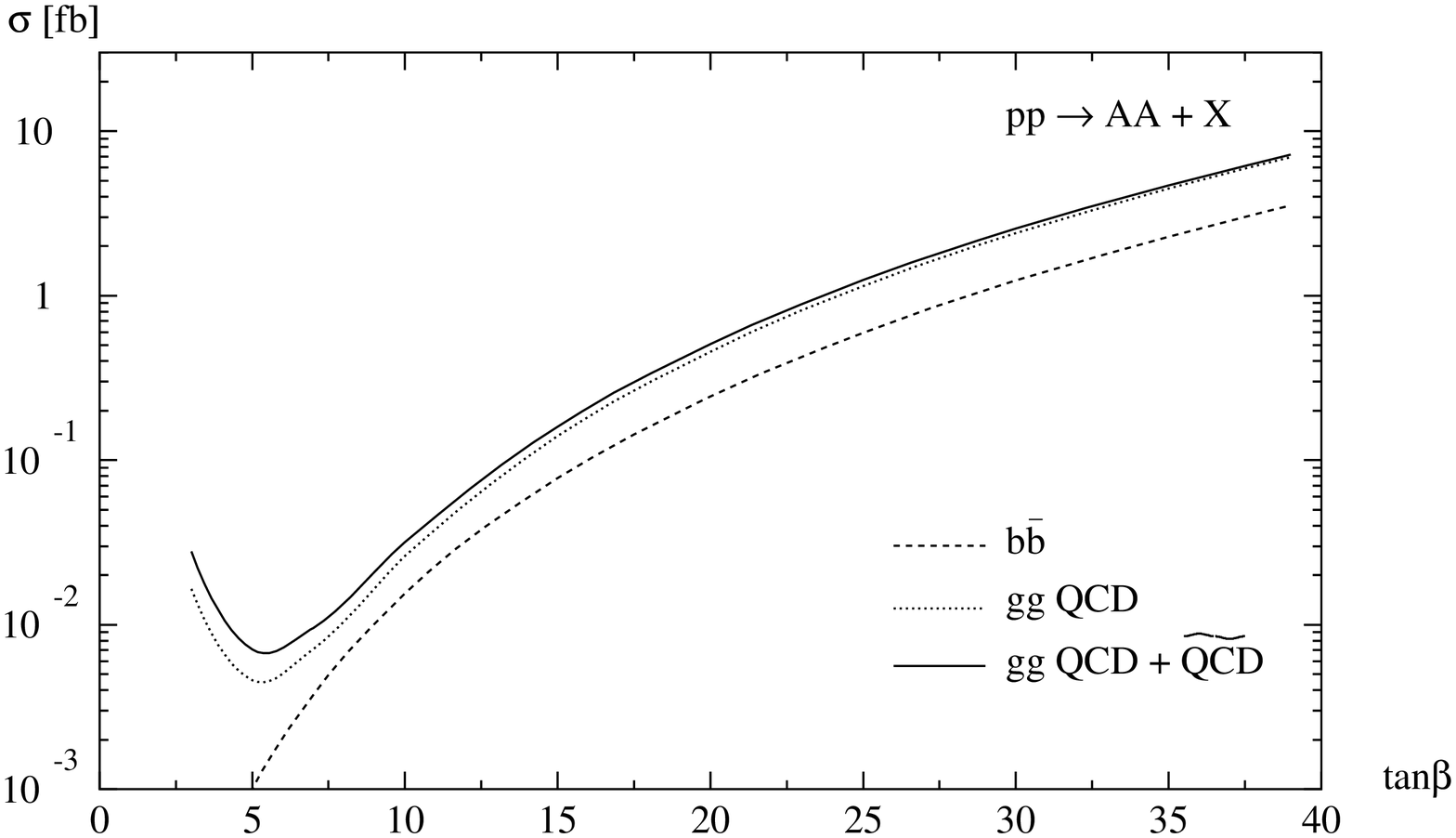,width=12cm}} \\
(b)
\end{tabular}
\caption{Total cross sections $\sigma$ (in fb) of $pp\to A^0A^0+X$ via
$b\bar b$ annihilation (dashed lines) and $gg$ fusion (solid lines) at the LHC
(a) as functions of $m_{A^0}$ for $\tan\beta=3$ (starting at 
$m_{A^0}=240$~GeV) and 30 (starting at $m_{A^0}=90$~GeV); and (b) as functions
of $\tan\beta$ for $m_{A^0}=300$~GeV.
For comparison, also the quark loop contribution to $gg$ fusion (dotted lines)
is shown.}
\label{fig:AA}
\end{center}
\end{figure}

\begin{figure}[ht]
\begin{center}
\begin{tabular}{c}
\parbox{12cm}{\epsfig{figure=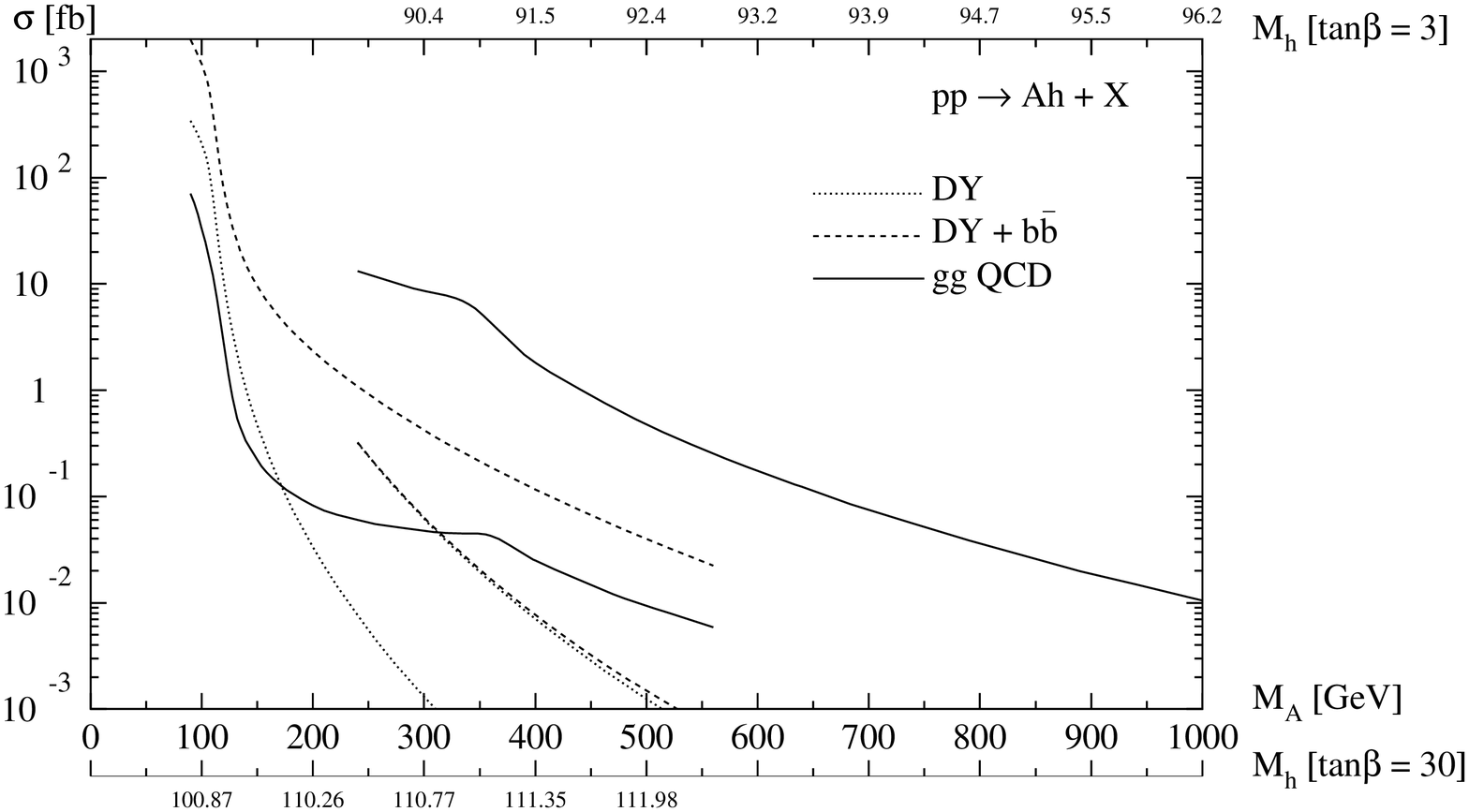,width=12cm}} \\
(a) \\
\parbox{12cm}{\epsfig{figure=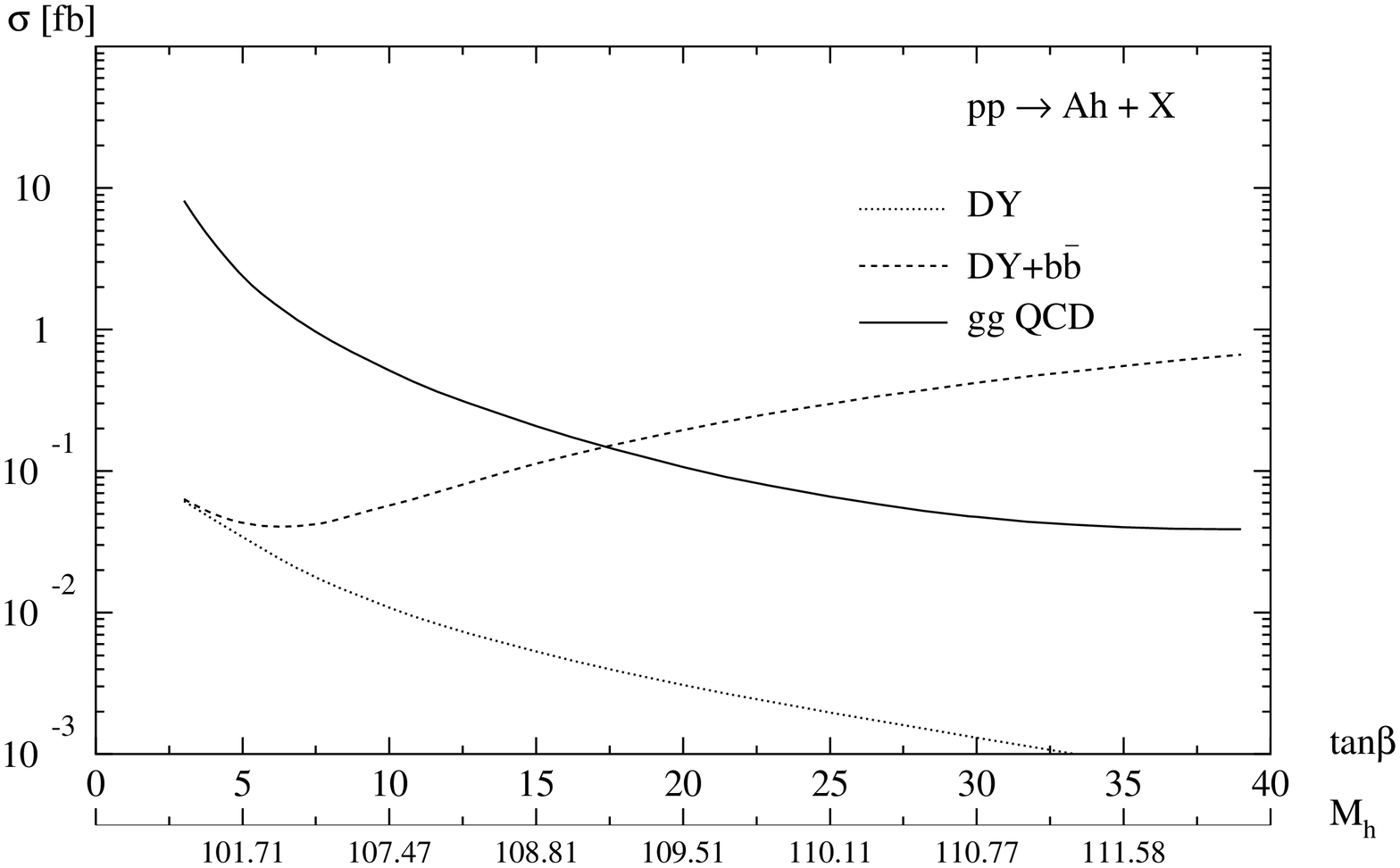,width=12cm}} \\
(b)
\end{tabular}
\caption{Total cross sections $\sigma$ (in fb) of $pp\to h^0A^0+X$ via
$q\bar q$ annihilation (dashed lines) and $gg$ fusion (solid lines) at the LHC
(a) as functions of $m_{A^0}$ for $\tan\beta=3$ (starting at
$m_{A^0}=240$~GeV) and 30 (starting at $m_{A^0}=90$~GeV); and (b) as functions
of $\tan\beta$ for $m_{A^0}=300$~GeV.
For comparison, also the Drell-Yan contribution to $q\bar q$ annihilation 
(dotted lines) is shown.}
\label{fig:hA}
\end{center}
\end{figure}

\begin{figure}[ht]
\begin{center}
\begin{tabular}{c}
\parbox{12cm}{\epsfig{figure=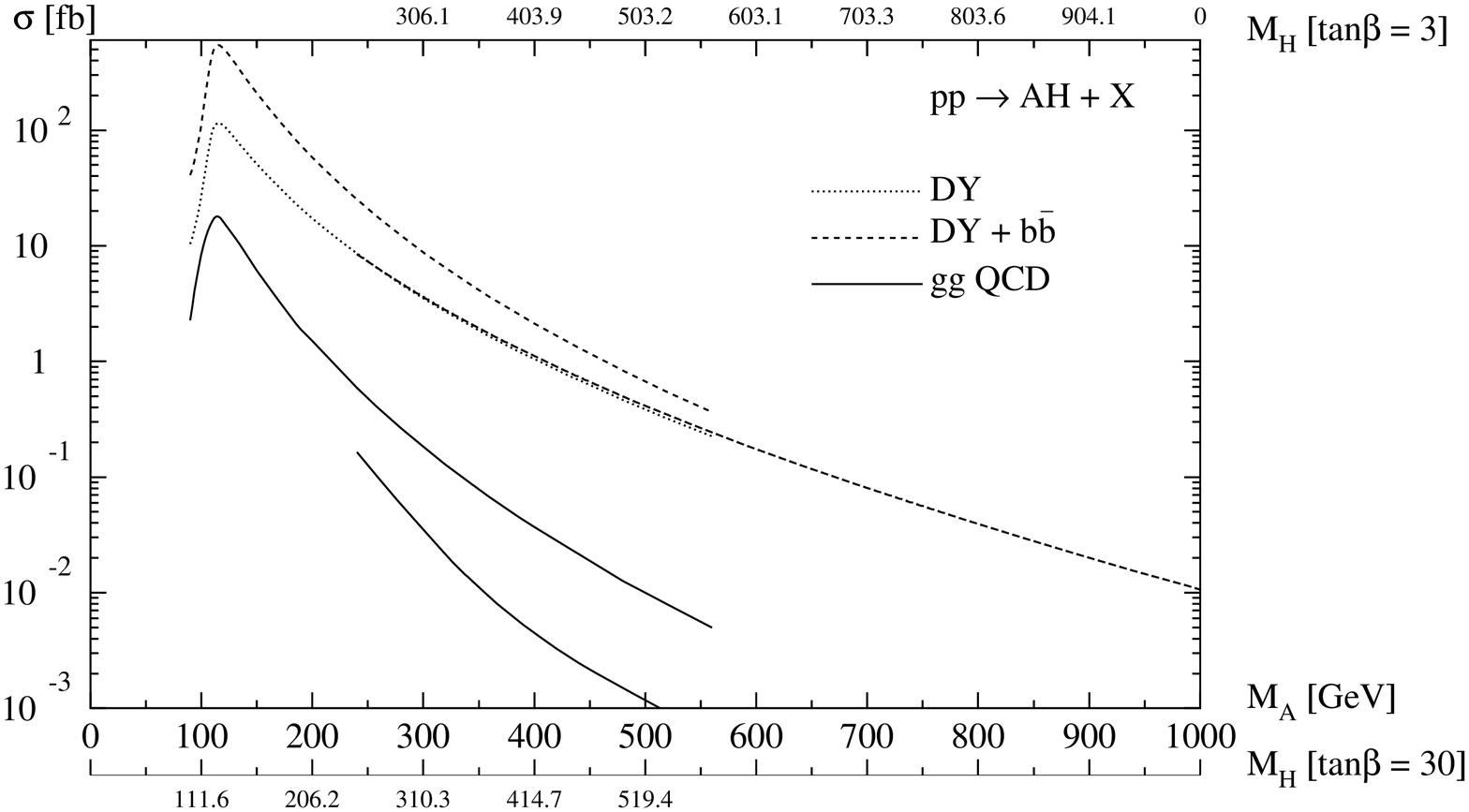,width=12cm}} \\
(a) \\
\parbox{12cm}{\epsfig{figure=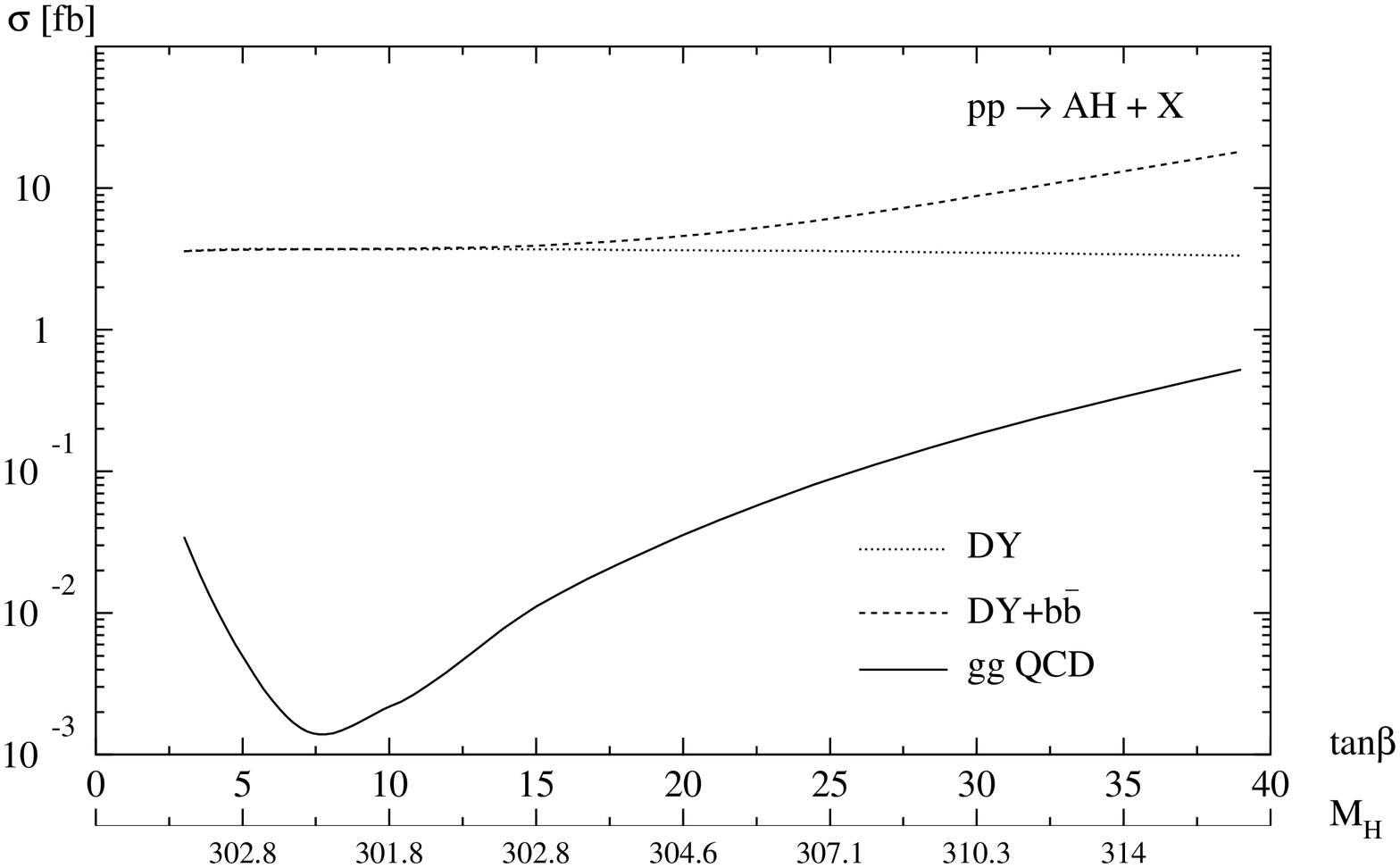,width=12cm}} \\
(b)
\end{tabular}
\caption{Total cross sections $\sigma$ (in fb) of $pp\to H^0A^0+X$ via
$q\bar q$ annihilation (dashed lines) and $gg$ fusion (solid lines) at the LHC
(a) as functions of $m_{A^0}$ for $\tan\beta=3$ (starting at
$m_{A^0}=240$~GeV) and 30 (starting at $m_{A^0}=90$~GeV); and (b) as functions
of $\tan\beta$ for $m_{A^0}=300$~GeV.
For comparison, also the Drell-Yan contribution to $q\bar q$ annihilation 
(dotted lines) is shown.}
\label{fig:HA}
\end{center}
\end{figure}

We now study the fully integrated cross sections of $pp\to\phi_1\phi_2+X$ at
the LHC, with c.m.\ energy $\sqrt S=14$~TeV.
Figures~\ref{fig:hh}--\ref{fig:HA} refer to the cases
$\phi_1\phi_2=h^0h^0,h^0H^0,H^0H^0,A^0A^0,$ $h^0A^0,H^0A^0$, respectively.
In part (a) of each figure, the $m_{A^0}$ dependence is studied for
$\tan\beta=3$ and 30 while, in part~(b), the $\tan\beta$ dependence is studied
for $m_{A^0}=300$~GeV.
We note that the SUGRA-inspired MSSM with our choice of input parameters does
not permit $\tan\beta$ and $m_{A^0}$ to be simultaneously small, due to the
experimental lower bound on the selectron mass \cite{ruh}.
This explains why the curves for $\tan\beta=3$ in 
Figs.~\ref{fig:hh}--\ref{fig:HA}(a) only start at $m_{A^0}\approx240$~GeV,
while those for $\tan\beta=30$ already start at $m_{A^0}\approx90$~GeV.
On the other hand, $\tan\beta$ and $m_{A^0}$ cannot be simultaneously large 
either, due to the experimental lower bounds on the chargino and neutralino
masses \cite{ruh}.
For this reason, the curves for $\tan\beta=30$ in
Figs.~\ref{fig:hh}--\ref{fig:HA}(a) already end at $m_{A^0}\approx560$~GeV.
Finally, the experimental $m_{h^0}$ lower bound \cite{ruh} enforces
$\tan\beta\agt3$ if $m_{A^0}=300$~GeV, which is reflected in
Figs.~\ref{fig:hh}--\ref{fig:HA}(b).

In Figs.~\ref{fig:hh}--\ref{fig:AA}, the $b\bar b$-annihilation contributions 
(dashed lines), which originate from Yukawa-enhanced amplitudes, and the total
$gg$-fusion contributions (solid lines), corresponding to the coherent
superposition of quark and squark loop amplitudes, are presented separately.
For a comparison with future experimental data, they should be added.
For comparison, also the $gg$-fusion contributions due to quark loops only
(dotted  lines) are shown.
We first assess the relative importance of the $b\bar b$-annihilation and
$gg$-fusion contributions.
In the case of $h^0h^0$ production, $b\bar b$ annihilation is more important
than $gg$ fusion for intermediate values of $m_{A^0}$, around 300~GeV, except
at the edges of the allowed $\tan\beta$ range, while it is greatly suppressed
for large values of $m_{A^0}$, independent of $\tan\beta$ (see
Fig.~\ref{fig:hh}).
In the case of $h^0H^0$ production, $b\bar b$ annihilation dominates for 
intermediate to large values of $m_{A^0}$ and large values of $\tan\beta$,
while it yields an insignificant contribution for small values of $\tan\beta$,
independent of $m_{A^0}$ (see Fig.~\ref{fig:hH}).
As for $H^0H^0$ and $A^0A^0$ production, $b\bar b$ annihilation is suppressed
compared to $gg$ fusion.
For $\tan\beta\agt8$, the suppression factor is modest, ranging between 2 and
3, but it dramatically increases as $\tan\beta$ becomes smaller (see
Figs.~\ref{fig:HH} and \ref{fig:AA}).
In order to avoid confusion, we should mention that the $b\bar b$-annihilation
contribution for $\tan\beta=3$ is too small to be visible in
Figs.~\ref{fig:hH}--\ref{fig:AA}(a).
We now investigate the size of the supersymmetric corrections to $gg$ fusion,
{\it i.e.}, the effect of including the squark loops.
We observe that these corrections can be of either sign and reach a magnitude 
of up to 90\%.
Specifically, for $h^0h^0$, $h^0H^0$, $H^0H^0$, and $A^0A^0$ production, they
vary within the ranges $-10\%$ to $+3\%$, $-16\%$ to $+32\%$, $-32\%$ to
$+24\%$, and $0\%$ to $+90\%$, respectively, for the values of $m_{A^0}$ and
$\tan\beta$ considered in Figs.~\ref{fig:hh}--\ref{fig:AA}(a).
The fact that they are relatively modest in most cases is characteristic for
our SUGRA-inspired MSSM scenario.
This is partly due to the destructive interference of quark and squark loop
amplitudes and to the suppression of the latter by heavy-squark propagators.
By contrast, in Ref.~\cite{bel}, the supersymmetric corrections were found
reach values in excess of 100.
Since the authors of Ref.~\cite{bel} did not specify all their input
parameters, we could not reproduce their numerical results for the squark loop
contributions.

In Figs.~\ref{fig:hA} and \ref{fig:HA}, the total $q\bar q$-annihilation
contributions (dashed lines), corresponding to the coherent superposition of
Drell-Yan and Yukawa-enhanced amplitudes, and the $gg$-fusion contributions
(solid lines), which now only receive contributions from quark loops, are
given separately.
For comparison, also the pure Drell-Yan contributions (dotted lines) are
shown.
Again, we first compare the total $q\bar q$-annihilation contributions with
the $gg$-fusion ones.
In the case of $h^0A^0$ production, $q\bar q$ annihilation dominates for large
values of $\tan\beta$, independent of $m_{A^0}$, while, at the lower end of
the allowed $\tan\beta$ range, it is suppressed by a  factor of 40 and more,
depending on the value of $m_{A^0}$.
On the other hand, in the case of $H^0A^0$ production, the
$q\bar q$-annihilation contribution always overshoots the $gg$-fusion one by
at least one order of magnitude.
We then examine the effect of including the Yukawa-enhanced amplitudes in the
evaluation of the $q\bar q$-annihilation cross section.
In the case of $h^0A^0$ production, there is a dramatic enhancement for large
values of $\tan\beta$, which may reach several orders of magnitude for large
values of $m_{A^0}$.
In the case of $H^0A^0$ production, there is also an enhancement for large
values of $\tan\beta$, but it is much more moderate, less than a factor of
three.
It is interesting to observe that the Drell-Yan cross section of $H^0A^0$ 
production is fairly independent of $\tan\beta$ unless $m_{A^0}$ is close to
its lower bound.
This may be understood by observing that $\sin(\alpha-\beta)$, which governs 
the $H^0A^0Z$ coupling $g_{H^0A^0Z}$, defined in Eq.~(\ref{eq:hAZ}), is then
always very close to $-1$.
This is also apparent from Fig.~2 of Ref.~\cite{spi}.

At this point, we should estimate the theoretical uncertainties in our
predictions.
As a typical example, we consider the cross section of $pp\to h^0h^0+X$ for
$\tan\beta=3$ and $m_{A^0}=300$~GeV.
In order to obtain a hint on the size of the as-yet unknown
next-to-leading-order (NLO) corrections, we define the renormalization and
factorization scales as $M=\mu_r=\xi\sqrt s$ and vary the scale parameter 
$\xi$ in the range $1/2<\xi<2$.
The resulting variation in cross section amounts to $\pm8\%$ 
in the case of $b\bar b$ annihilation and to $\pm11\%$ in the case of $gg$ 
fusion.
At NLO, one also needs to specify a renormalization scheme for the definition
of the $b$-quark mass, which enters our analysis through the $b$-quark Yukawa
coupling.
Our LO analysis is appropriate for the on-mass-shell scheme, which uses pole
masses as basic parameters.
The modified minimal-subtraction ($\overline{\rm MS}$) scheme \cite{bar} 
provides a popular alternative.
For example, a pole mass of $m_b=4.6$~GeV \cite{pdg} corresponds to an
$\overline{\rm MS}$ mass of $\overline{m}_b^{(5)}(\mu_r)=2.7$~GeV for the
typical choice of renormalization scale $\mu_r=\sqrt s=300$~GeV.
Recalling that the leading behaviour of Eq.~(\ref{eq:hh}) in $m_b$ is 
quadratic, switching to the $\overline{\rm MS}$ scheme would thus, at first
sight, lead to a suppression of the cross section by a factor of approximately
1/3.
However, we must keep in mind that this reduction should be largely cancelled,
up to terms that are formally beyond NLO, by a respective shift in the NLO
correction. 
Another source of uncertainty is related to the choice of PDF's.
In fact, there exist significant differences in the extraction of the 
$b$-quark PDF among different PDF sets, which are related to the threshold
treatment of the $g\to b\bar b$ splitting, the choice of the $b$-quark mass,
the dependence of the evolution on the latter, {\it etc.}
In the case of a LO analysis with $n_f=5$ massless quark favours, which is
considered here, the use of CTEQ5L \cite{lai} together with the scale choice
$M=\mu_r=\sqrt s$ should be appropriate in the sense that these issues can
largely be bypassed.
If we employ the LO PDF set by Martin, Roberts, Stirling, and Thorne (MRST)
\cite{mar}, with $\Lambda_{\rm QCD}^{(5)}=132$~MeV, then the
$b\bar b$-annihilation and $gg$-fusion cross sections increase by $10\%$ and
$4\%$, respectively, relative to their default values.

\section{\label{sec:four}Conclusions}

We analytically calculated the cross sections of the partonic subprocesses
$q\bar q\to\phi_1\phi_2$ and $gg\to\phi_1\phi_2$, where $\phi_i=h^0,H^0,A^0$,
to LO in the MSSM.
We included the Drell-Yan and Yukawa-enhanced contributions to $q\bar q$ 
annihilation (see Fig.~\ref{fig:tree}) and the quark and squark loop
contributions to $gg$ fusion (see Fig.~\ref{fig:loop}).
We listed our formulas for the $q\bar q$-annihilation cross sections and the
squark loop amplitudes, for which we found rather compact expressions.
As for the quark loop contributions, we found complete agreement with 
Ref.~\cite{ple}.

We then quantitatively investigated the inclusive cross sections of
$pp\to\phi_1\phi_2+X$ at the LHC adopting a favorable SUGRA-inspired MSSM
scenario, varying the input parameters $m_{A^0}$ and $\tan\beta$.
The results are presented in Figs.~\ref{fig:hh}--\ref{fig:HA}.
We found that the Yukawa-enhanced $q\bar q$-annihilation contribution, which 
had previously been neglected, can play a leading role, especially for
$h^0h^0$ production if $m_{A^0}$ is of order 300~GeV and for $h^0H^0$,
$h^0A^0$, and $H^0A^0$ production if $\tan\beta$ is large.
The supersymmetric corrections to $gg$ fusion, which are present for $h^0h^0$,
$h^0H^0$, $H^0H^0$, and $A^0A^0$ production, can be of either sign and reach
a magnitude of up to 90\%.
Our numerical results for these corrections disagree with those presented in
Ref.~\cite{bel}.
For each process $pp\to\phi_1\phi_2+X$, the combined cross section,
{\it i.e.}, the sum of the full $q\bar q$-annihilation and $gg$-fusion 
contributions, varies by several orders of magnitude as the values of
$m_{A^0}$ and $\tan\beta$ are changed within their allowed ranges, and its
maximum value is typically between $10^2$~fb$^{-1}$ and $10^3$~fb$^{-1}$.
If we assume the integrated luminosity per year to be at its design value of
$L=100$~fb$^{-1}$ for each of the two LHC experiments, ATLAS and CMS, then 
this translates into a maximum of 20.000 to 200.000 events per year for each
of these signal processes.

A comprehensive discussion of the background processes competing with the
$pp\to\phi_1\phi_2+X$ signals at the LHC lies beyond the scope of our study.
However, we should briefly mention them and quote the relevant literature.
Without specifying the decay channels of the $\phi_i$ bosons, one expects the
major backgrounds to arise from the pair production of neutral gauge bosons,
the associate production of a neutral gauge boson and a neutral Higgs boson,
and the continuum production of the respective $\phi_1\phi_2$ decay products.
Published signal-to-background analyses \cite{dai} have concentrated on the
$\phi_1\phi_2\to b\bar bb\bar b$ signals and their irreducible continuum
backgrounds, due to the partonic subprocesses $gg,q\bar q\to b\bar bb\bar b$,
which are dominantly of pure QCD origin.
It has been demonstrated that, after optimizing the acceptance cuts, the LHC
experiments might discover a signal, with experimental significance in excess
of 5, if $\tan\beta\alt3$ or $\tan\beta\agt50$.

\vspace{1cm}
\noindent
{\bf Acknowledgements}
\smallskip

We thank S. Moretti for drawing our attention to Ref.~\cite{dai}.
The work of A.A.B.B. was supported by the Deutsches Elektronen-Synchrotron
DESY.
The work of B.A.K. was supported in part by the Deutsche
Forschungsgemeinschaft through Grant No.\ KN~365/1-1, by the
Bundesministerium f\"ur Bildung und Forschung through Grant No.\ 05~HT9GUA~3,
and by the European Commission through the Research Training Network
{\it Quantum Chromodynamics and the Deep Structure of Elementary Particles}
under Contract No.\ ERBFMRX-CT98-0194.

\def\theequation{\Alph{section}.\arabic{equation}}
\begin{appendix}
\setcounter{equation}{0}
\section{Relevant Higgs and squark couplings}

In this appendix, we list the trilinear self-couplings of the $h^0$, $H^0$, 
and $A^0$ bosons as well as their couplings to the $Z$ boson and the $t$ and
$b$ quarks.
Furthermore, we collect the couplings of these Higgs bosons to the squarks
$\tilde q_i$, with $q=t,b$ and $i=1,2$, which are not contained in Appendix~A
of Ref.~\cite{hh}.
For convenience, we introduce the short-hand notations
$s_\alpha=\sin\alpha$, $c_\alpha=\cos\alpha$,
$s_\beta=\sin\beta$, $c_\beta=\cos\beta$,
$s_{2\beta}=\sin(2\beta)$, $c_{2\beta}=\cos(2\beta)$,
$s_\pm=\sin(\alpha\pm\beta)$, and $c_\pm=\cos(\alpha\pm\beta)$.

The trilinear self-couplings of the $h^0$, $H^0$, and $A^0$ bosons are given 
by \cite{hab}
\begin{eqnarray}
g_{h^0h^0h^0}&=&-\frac{3m_Z}{2c_w}c_{2\alpha}s_+,\qquad
g_{h^0h^0H^0}=-\frac{m_Z}{2c_w}(2s_{2\alpha}s_+-c_{2\alpha}c_+),\nonumber\\
g_{h^0H^0H^0}&=&\frac{m_Z}{2c_w}(2s_{2\alpha}c_++c_{2\alpha}s_+),\qquad
g_{H^0H^0H^0}=-\frac{3m_Z}{2c_w}c_{2\alpha}c_+,\nonumber\\
g_{h^0A^0A^0}&=&-\frac{m_Z}{2c_w}c_{2\beta}s_+,\qquad
g_{H^0A^0A^0}=\frac{m_Z}{2c_w}c_{2\beta}c_+.
\label{eq:hhh}
\end{eqnarray}
Their couplings to the $Z$ boson are given by \cite{hab}
\begin{equation}
g_{h^0A^0Z}=\frac{c_-}{2c_w},\qquad
g_{H^0A^0Z}=\frac{s_-}{2c_w}.
\label{eq:hAZ}
\end{equation}
Their couplings to the $t$ and $b$ quarks are given by \cite{hab}
\begin{eqnarray}
g_{h^0tt}&=&-\frac{m_tc_\alpha}{2m_Ws_\beta},\qquad
g_{h^0bb}=\frac{m_bs_\alpha}{2m_Wc_\beta},\nonumber\\
g_{H^0tt}&=&-\frac{m_ts_\alpha}{2m_Ws_\beta},\qquad
g_{H^0bb}=-\frac{m_bc_\alpha}{2m_Wc_\beta},\nonumber\\
g_{A^0tt}&=&-\frac{m_t\cot\beta}{2m_W},\qquad
g_{A^0bb}=-\frac{m_b\tan\beta}{2m_W}.
\label{eq:hqq}
\end{eqnarray}
The missing couplings of these Higgs bosons to the squarks are given by
\cite{hab}
\begin{eqnarray}
\left(
\begin{array}{cc}
g_{A^0\tilde t_1\tilde t_1} & g_{A^0\tilde t_1\tilde t_2} \\
g_{A^0\tilde t_2\tilde t_1} & g_{A^0\tilde t_2\tilde t_2} \\
\end{array}
\right)
&=&\frac{m_t(\mu+A_t\cot\beta)}{2m_W}\left(
\begin{array}{cc}
0 & 1 \\
-1 & 0 \\
\end{array}
\right),
\nonumber\\
\left(
\begin{array}{cc}
g_{A^0\tilde b_1\tilde b_1} & g_{A^0\tilde b_1\tilde b_2} \\
g_{A^0\tilde b_2\tilde b_1} & g_{A^0\tilde b_2\tilde b_2} \\
\end{array}
\right) 
&=&\frac{m_b(\mu+A_b\tan\beta)}{2m_W}\left(
\begin{array}{cc}
0 & 1 \\
-1 & 0 \\
\end{array}
\right),
\nonumber\\
\left(
\begin{array}{cc}
g_{h^0h^0\tilde t_1\tilde t_1} & g_{h^0h^0\tilde t_1\tilde t_2} \\
g_{h^0h^0\tilde t_2\tilde t_1} & g_{h^0h^0\tilde t_2\tilde t_2} \\
\end{array}
\right) 
&=&{\cal M}^{\tilde t}\left(
\begin{array}{cc}
\frac{c_{2\alpha}\left(I_t-s_w^2Q_t\right)}{2c_w^2}
-\frac{m_t^2c_{\alpha}^2}{2m_W^2s_{\beta}^2} & 0 \\
0 & \frac{c_{2\alpha}s_w^2Q_t}{2c_w^2}
-\frac{m_t^2c_{\alpha}^2}{2m_W^2s_{\beta}^2} \\
\end{array}
\right)\left({\cal M}^{\tilde t}\right)^T,
\nonumber\\
\left(
\begin{array}{cc}
g_{h^0h^0\tilde b_1\tilde b_1} & g_{h^0h^0\tilde b_1\tilde b_2} \\
g_{h^0h^0\tilde b_2\tilde b_1} & g_{h^0h^0\tilde b_2\tilde b_2} \\
\end{array}
\right) 
&=&{\cal M}^{\tilde b}\left(
\begin{array}{cc}
\frac{c_{2\alpha}(I_b-s_w^2Q_b)}{2c_w^2}
-\frac{m_b^2s_{\alpha}^2}{2m_W^2c_{\beta}^2} & 0 \\
0 & \frac{c_{2\alpha}s_w^2Q_b}{2c_w^2}
-\frac{m_b^2s_{\alpha}^2}{2m_W^2c_{\beta}^2} \\
\end{array}
\right)\left({\cal M}^{\tilde b}\right)^T,
\nonumber\\
\left(
\begin{array}{cc}
g_{h^0H^0\tilde t_1\tilde t_1} & g_{h^0H^0\tilde t_1\tilde t_2} \\
g_{h^0H^0\tilde t_2\tilde t_1} & g_{h^0H^0\tilde t_2\tilde t_2} \\
\end{array}
\right)
&=&{\cal M}^{\tilde t}\left(
\begin{array}{cc}
\frac{s_{2\alpha}\left(I_t-s_w^2Q_t\right)}{2c_w^2}
-\frac{m_t^2s_{2\alpha}}{4m_W^2s_{\beta}^2} & 0 \\
0 & \frac{s_{2\alpha}s_w^2Q_t}{2c_w^2}
-\frac{m_t^2s_{2\alpha}}{4m_W^2s_{\beta}^2} \\
\end{array}
\right)\left({\cal M}^{\tilde t}\right)^T,
\nonumber\\
\left(
\begin{array}{cc}
g_{h^0H^0\tilde b_1\tilde b_1} & g_{h^0H^0\tilde b_1\tilde b_2} \\
g_{h^0H^0\tilde b_2\tilde b_1} & g_{h^0H^0\tilde b_2\tilde b_2} \\
\end{array}
\right) 
&=&{\cal M}^{\tilde b}\left(
\begin{array}{cc}
\frac{s_{2\alpha}(I_b-s_w^2Q_b)}{2c_w^2}
+\frac{m_b^2s_{2\alpha}}{4m_W^2c_{\beta}^2} & 0 \\
0 & \frac{s_{2\alpha}s_w^2Q_b}{2c_w^2}
+\frac{m_b^2s_{2\alpha}}{4m_W^2c_{\beta}^2} \\
\end{array}
\right)\left({\cal M}^{\tilde b}\right)^T,
\nonumber\\
\left(
\begin{array}{cc}
g_{H^0H^0\tilde t_1\tilde t_1} & g_{H^0H^0\tilde t_1\tilde t_2} \\
g_{H^0H^0\tilde t_2\tilde t_1} & g_{H^0H^0\tilde t_2\tilde t_2} \\
\end{array}
\right) 
&=&{\cal M}^{\tilde t}\left(
\begin{array}{cc}
-\frac{c_{2\alpha}\left(I_t-s_w^2Q_t\right)}{2c_w^2}
-\frac{m_t^2s_{\alpha}^2}{2m_W^2s_{\beta}^2} & 0 \\
0 & -\frac{c_{2\alpha}s_w^2Q_t}{2c_w^2}
-\frac{m_t^2s_{\alpha}^2}{2m_W^2s_{\beta}^2} \\
\end{array}
\right)\left({\cal M}^{\tilde t}\right)^T,
\nonumber\\
\left(
\begin{array}{cc}
g_{H^0H^0\tilde b_1\tilde b_1} & g_{H^0H^0\tilde b_1\tilde b_2} \\
g_{H^0H^0\tilde b_2\tilde b_1} & g_{H^0H^0\tilde b_2\tilde b_2} \\
\end{array}
\right) 
&=&{\cal M}^{\tilde b}\left(
\begin{array}{cc}
-\frac{c_{2\alpha}(I_b-s_w^2Q_b)}{2c_w^2}
-\frac{m_b^2c_{\alpha}^2}{2m_W^2c_{\beta}^2} & 0 \\
0 & -\frac{c_{2\alpha}s_w^2Q_b}{2c_w^2}
-\frac{m_b^2c_{\alpha}^2}{2m_W^2c_{\beta}^2} \\
\end{array}
\right)\left({\cal M}^{\tilde b}\right)^T,
\nonumber\\
\left(
\begin{array}{cc}
g_{A^0A^0\tilde t_1\tilde t_1} & g_{A^0A^0\tilde t_1\tilde t_2} \\
g_{A^0A^0\tilde t_2\tilde t_1} & g_{A^0A^0\tilde t_2\tilde t_2} \\
\end{array}
\right)
&=&{\cal M}^{\tilde t}\left(
\begin{array}{cc}
\frac{c_{2\beta}\left(I_t-s_w^2Q_t\right)}{2c_w^2}
-\frac{m_t^2\cot^2\beta}{2m_W^2} & 0 \\
0 & \frac{c_{2\beta}s_w^2Q_t}{2c_w^2}-\frac{m_t^2\cot^2\beta}{2m_W^2} \\
\end{array}
\right)\left({\cal M}^{\tilde t}\right)^T,
\nonumber\\
\left(
\begin{array}{cc}
g_{A^0A^0\tilde b_1\tilde b_1} & g_{A^0A^0\tilde b_1\tilde b_2} \\
g_{A^0A^0\tilde b_2\tilde b_1} & g_{A^0A^0\tilde b_2\tilde b_2} \\
\end{array}
\right) 
&=&{\cal M}^{\tilde b}\left(
\begin{array}{cc}
\frac{c_{2\beta}(I_b-s_w^2Q_b)}{2c_w^2}-\frac{m_b^2\tan^2\beta}{2m_W^2} &
0 \\
0 & \frac{c_{2\beta}s_w^2Q_b}{2c_w^2}-\frac{m_b^2\tan^2\beta}{2m_W^2} \\
\end{array}
\right)\left({\cal M}^{\tilde b}\right)^T.
\label{eq:hss}
\end{eqnarray}
Here, ${\cal M}^{\tilde q}$ denotes the mixing matrix which rotates the left-
and right-handed squark fields, $\tilde q_L$ and $\tilde q_R$, into the mass
eigenstates $\tilde q_i$.
Its definition may be found in Eq.~(A.1) of Ref.~\cite{hh}.
Relations similar to Eq.~(\ref{eq:hss}) are valid for the squarks of the first
and second generations, which are also included in our analysis.
However, in these cases, we neglect terms which are suppressed by the
smallness of the corresponding light-quark masses.

\setcounter{equation}{0}
\section{Squark loop form factors}

In this appendix, we express the squark triangle and box form factors,
$\tilde F_\triangle$, $\tilde F_\Box$, and $\tilde G_\Box$, in terms of the 
standard scalar three- and four-point functions, which we abbreviate as
$C_{ijk}^{ab}(c)=C_0\left(a,b,c,m_{\tilde q_i}^2,m_{\tilde q_j}^2,
m_{\tilde q_k}^2\right)$ and
$D_{ijkl}^{abcd}(e,f)=D_0\left(a,b,c,d,e,f,m_{\tilde q_i}^2,m_{\tilde q_j}^2,
m_{\tilde q_k}^2,m_{\tilde q_l}^2\right)$, respectively.
The definitions of the latter may be found in Eq.~(5) of Ref.~\cite{wh1}.

We have
\begin{eqnarray}
\tilde F_{\triangle}&=&\sum_{\tilde q}\sum_{i=1}^2
\left(g_{\phi_1\phi_2h^0}g_{h^0\tilde q_i\tilde q_i}{\cal P}_{h^0}(s)
+g_{\phi_1\phi_2H^0}g_{H^0\tilde q_i\tilde q_i}{\cal P}_{H^0}(s) 
-g_{\phi_1\phi_2\tilde q_i\tilde q_i}\right)F_1\left(s,m_{\tilde q_i}^2\right),
\nonumber\\
\tilde F_{\Box}&=&\frac{2}{s}\sum_{\tilde q}\sum_{i,j=1}^2
g_{\phi_1\tilde q_i\tilde q_j}g_{\phi_2\tilde q_j\tilde q_i}
F_2\left(s,t,h_1,h_2,m_{\tilde q_i}^2,m_{\tilde q_j}^2\right),
\nonumber\\
\tilde G_{\Box}&=&\frac{2}{sp_T^2}\sum_{\tilde q}\sum_{i,j=1}^2
g_{\phi_1\tilde q_i\tilde q_j}g_{\phi_2\tilde q_j\tilde q_i}
F_3\left(s,t,h_1,h_2,m_{\tilde q_i}^2,m_{\tilde q_j}^2\right).
\end{eqnarray}
Here, we have introduced the auxiliary functions
\begin{eqnarray}
F_1\left(s,m_{\tilde q_i}^2\right)
&=&2+4m_{\tilde q_i}^2C^{00}_{iii}(s),
\nonumber\\
F_2\left(s,t,h_1,h_2,m_{\tilde q_i}^2,m_{\tilde q_j}^2\right)
&=&-t_1C^{h_10}_{ijj}(t)-t_2C^{h_20}_{ijj}(t) 
+2sm_{\tilde q_i}^2D^{h_1h_200}_{ijii}(s,t)
+s\left(\frac{p_T^2}{2}+m_{\tilde q_i}^2\right)
\nonumber\\
&&{}\times D^{h_10h_20}_{ijji}(t,u)
+(t\leftrightarrow u),
\nonumber\\
F_3\left(s,t,h_1,h_2,m_{\tilde q_i}^2,m_{\tilde q_j}^2\right)
&=&-s\left(t+m_{\tilde q_i}^2\right)C^{000}_{iii}(s)
+sm_{\tilde q_i}^2C^{000}_{jjj}(s)
-tt_1C^{h_10}_{ijj}(t)
-tt_2C^{h_20}_{ijj}(t)
\nonumber\\
&&{}+(t^2-h_1h_2)C^{h_1h_2}_{iji}(s)
+\left[st^2-2t_1t_2m_{\tilde q_i}^2
+2sm_{\tilde q_i}^2\left(m_{\tilde q_i}^2-m_{\tilde q_j}^2\right)\right]
\nonumber\\
&&{}\times D^{h_1h_200}_{ijii}(s,t)
-2stm_{\tilde q_i}^2D^{h_1h_200}_{jijj}(s,t)
+sm_{\tilde q_i}^2\left(p_T^2+m_{\tilde q_i}^2-m_{\tilde q_j}^2\right)
\nonumber\\
&&{}\times D^{h_10h_20}_{ijji}(t,u)+(t\leftrightarrow u),
\end{eqnarray}
where $t_i=t-h_i$.

\end{appendix}

\newpage

\newpage

\newpage

\newpage

\newpage

\newpage

\newpage

\newpage


\begin{thebibliography}{99}

\bibitem{kun} Z. Kunszt and F. Zwirner,
Nucl.\ Phys.\ {\bf B385}, 3 (1992), and references cited therein.

\bibitem{hab} H. E. Haber and G. L. Kane,
Phys.\ Rep.\ {\bf117}, 75 (1985);
J. F. Gunion and H. E. Haber,
Nucl.\ Phys.\ {\bf B272}, 1 (1986); {\bf B402}, 567(E) (1993);
{\bf B278}, 449 (1986); {\bf B402}, 569(E) (1993);
{\bf B307}, 445 (1988); {\bf B402}, 569(E) (1993);
J. F. Gunion, H. E. Haber, G. Kane, and S. Dawson,
{\it The Higgs Hunter's Guide} (Addison-Wesley, Redwood City, 1990).

\bibitem{dic} D. A. Dicus, C. Kao, and S. S. D. Willenbrock,
Phys.\ Lett.\ B {\bf203}, 457 (1988);
E. W. N. Glover and J. J. van der Bij,
Nucl.\ Phys.\ {\bf B309}, 282 (1988).

\bibitem{eic} E. Eichten, I. Hinchliffe, K. Lane, and C. Quigg,
Rev.\ Mod.\ Phys.\ {\bf56}, 579 (1984); {\bf58}, 1065(E) (1986);
N. G. Deshpande, X. Tata, and D. A. Dicus,
Phys.\ Rev.\ D {\bf29}, 1527 (1984).

\bibitem{hh} A. A. Barrientos Bendez\'u and B. A. Kniehl,
Nucl.\ Phys.\ {\bf B568}, 305 (2000).

\bibitem{wil} S. S. D. Willenbrock,
Phys.\ Rev.\ D {\bf35}, 173 (1987);
J. Yi, H. Liang, M. Wen-Gan, Y. Zeng-Hui, and H. Meng,
J. Phys.\ G {\bf23}, 385 (1997); {\bf23}, 1151(E) (1997);
J. Yi, M. Wen-Gan, H. Liang, H. Meng, and Y. Zeng-Hui,
J. Phys.\ G {\bf24}, 83 (1998);
A. Krause, T. Plehn, M. Spira, and P. M. Zerwas,
Nucl.\ Phys.\ {\bf B519}, 85 (1998).

\bibitem{bre} O. Brein and W. Hollik,
Eur.\ Phys.\ J. C {\bf13}, 175 (2000).

\bibitem{daw} S. Dawson, S. Dittmaier, and M. Spira,
Phys.\ Rev.\ D {\bf58}, 115012 (1998).

\bibitem{ple} T. Plehn, M. Spira, and P. M. Zerwas,
Nucl.\ Phys.\ {\bf B479}, 46 (1996); {\bf B531}, 655(E) (1998).

\bibitem{bel} A. Belyaev, M. Drees, O. J. P. \'Eboli, J. K. Mizukoshi, and S. 
F. Novaes,
Phys.\ Rev.\ D {\bf60}, 075008 (1999);
M. Drees, private communication.

\bibitem{kil} A. Djouadi, W. Kilian, M. M\"uhlleitner, and P. M. Zerwas,
Eur.\ Phys.\ J. C {\bf10}, 45 (1999).

\bibitem{gun} J. F. Gunion, H. E. Haber, F. E. Paige, W.-K. Tung, and S. S. D. 
Willenbrock,
Nucl.\ Phys.\ {\bf B294}, 621 (1987);
R. M. Barnett, H. E. Haber, and D. E. Soper,
Nucl.\ Phys.\ {\bf B306}, 697 (1988);
F. I. Olness and W.-K. Tung,
Nucl.\ Phys.\ {\bf B308}, 813 (1988);
D. A. Dicus and S. Willenbrock,
Phys.\ Rev.\ D {\bf39}, 751 (1989);
D. A. Dicus and C. Kao,
Phys.\ Rev.\ D {\bf41}, 832 (1990);
V. Barger, R. J. N. Phillips, and D. P. Roy,
Phys.\ Lett.\ B {\bf324}, 236 (1994).

\bibitem{kal} A. Djouadi, J. Kalinowski, P. Ohmann, and P. M. Zerwas,
Z. Phys.\ C {\bf74}, 93 (1997), and references cited therein.

\bibitem{hem} R. Hempfling and B. Kniehl,
Z. Phys.\ C {\bf59}, 263 (1993).

\bibitem{wh} A. A. Barrientos Bendez\'u and B. A. Kniehl,
Phys.\ Rev.\ D {\bf59}, 015009 (1998).

\bibitem{pdg} Particle Data Group, D. E. Groom {\it et al.},
Eur.\ Phys.\ J. C {\bf15}, 1 (2000).

\bibitem{lai} CTEQ Collaboration, H. L. Lai {\it et al.},
Eur.\ Phys.\ J. C {\bf12}, 375 (2000).

\bibitem{djo} A. Djouadi, J.-L. Kneur, and G. Moultaka,
Report No.\ PM/98-27 and GDR-S-017 (1998).

\bibitem{ruh} V. Ruhlmann-Kleider,
in proceedings of XIX International Symposium on Lepton and Photon
Interactions at High Energies (Lepton-Photon 99), Stanford, California,
9--14 August 1999, edited by J. Jaros and M. Peskin (World Scientific, 
Singapore, 2000), p.~416.

\bibitem{spi} M. Spira,
Fortschr.\ Phys.\ {\bf46}, 203 (1998).

\bibitem{bar} W. A. Bardeen, A. J. Buras, D. W. Duke, and T. Muta,
Phys.\ Rev.\ D {\bf18}, 3998 (1978).

\bibitem{mar} A. D. Martin, R. G. Roberts, W. J. Stirling, and R. S. Thorne,
Phys.\ Lett.\ B {\bf443}, 301 (1998).

\bibitem{dai} J. Dai, J. F. Gunion, and R. Vega,
Phys.\ Lett.\ B {\bf371}, 71 (1996); {\bf387}, 801 (1996);
E. Richter-W\c as and D. Froidevaux,
Z. Phys.\ C {\bf76}, 665 (1997);
ATLAS Collaboration, A. Airapetian {\it et al.},
ATLAS Detector and Physics Performance: Technical Design Report, Vol.~II,
Report No.\ CERN/LHCC/99-15 and ATLAS TDR 15 (25 May 1999), p.~754;
A. Belyaev, M. Drees, and J. K. Mizukoshi,
Eur.\ Phys.\ J.\ C {\bf17}, 337 (2000);
R. Lafaye, D. J. Miller, M. M\"uhlleitner, and S. Moretti,
Report No.\ DESY 99-192, RAL-TR-99-083, and hep-ph/0002238, to appear in
{\it Proceedings of the Workshop on Physics at TeV Colliders}, Les Houches,
France, 7--18 June 1999.

\bibitem{wh1} A. A. Barrientos Bendez\'u and B. A. Kniehl,
Phys.\ Rev.\ D {\bf61}, 097701 (2000).

\end{thebibliography}
\end{document}